\documentclass[journal, onecolumn]{IEEEtran}
\usepackage{graphicx}
\usepackage{subfigure}
\usepackage{float}
\usepackage{amsmath}
\usepackage{epstopdf}
\usepackage{cases}
\usepackage{color}
\usepackage{algorithm}  
\usepackage{algpseudocode}  
\usepackage{amsmath}  

\ifCLASSINFOpdf

\else

\fi

\usepackage{caption}
\usepackage{mathtools}

\begin{document}

	\title{Improve Radar Sensing Performance of Multiple Roadside Units Cooperation via Space Registration}

	\author{
		Wangjun Jiang,~\IEEEmembership{Student Member,~IEEE,}
		Zhiqing Wei,~\IEEEmembership{ Member,~IEEE,} \\
		Bin Li,~\IEEEmembership{ Member,~IEEE,} 
		Zhiyong Feng,~\IEEEmembership{Senior Member,~IEEE,} 
		Zixi Fang,~\IEEEmembership{Student Member,~IEEE}
		\\
		\thanks{
			
			This work was supported in part by the National Key Research and Development Program under Grant 2020YFA0711302, and the BUPT Excellent Ph.D. Students Foundation.
		
			The authors are with Beijing University of Posts and Telecommunications, Beijing, China 100876 (email: \{jiangwangjun, weizhiqing, Binli, fengzy, fangzx\}@bupt.edu.cn).
			\emph{Corresponding authors: Zhiyong Feng and Zhiqing Wei.}}
		
	}

	\maketitle

	\begin{abstract}
	Roadside units (RSUs) can help vehicles sense the traffic environment, so as to improve traffic safety. Since the sensing capability of single RSU is limited, we propose a multiple RSUs cooperative radar sensing network (RSU-CRSN) with signal-level fusion technique. Spatial registration is an essential prerequisite and foundation for RSU-CRSN with signal-level fusion. In this paper, we present an adjustable beam enabled spatial registration algorithm (AB-SRA) that makes the sensing area of each RSU coincide by adjusting the sensing beam width of RSU. To adjust the width of sensing beam flexibly, a beamwidth adjustable beamforming algorithm (BABA) is proposed in this paper. Simulation results show that the performance of AB-SRA is close to perfect spatial registration. 
	
	\end{abstract}
	
	\begin{IEEEkeywords}
		Roadside unit, signal-level fusion, cooperative radar sensing, spatial registration, beamforming algorithm.
		
	\end{IEEEkeywords}
	
	\IEEEpeerreviewmaketitle
	
	\section{Introduction}\label{sec:introduction}

	Roadside units (RSUs) installed on the roadside can assist vehicles to sense the traffic environment, so as to improve traffic safety. The application of RSU in the intelligent transportation system (ITS) has been widely recognized \cite{[RSU_2020]}. 
	However, the sensing performance of single RSU is limited. To improve the sensing performance, a multiple RSUs cooperative radar sensing network (RSU-CRSN) with signal-level fusion scheme is proposed in this paper.
	Data fusion is critical for cooperative radar sensing network \cite{[information_fusion]}. There are three levels of data fusion algorithms, namely, tracking-level fusion, decision-level fusion and signal-level fusion \cite{[collaborative_sensing]}. 
	
	In a cooperative radar sensing network with tracking-level or decision-level fusion, RSUs firstly obtain the sensing information through radar signal processing, then the sensing information of multiple RSUs are fused to improve the sensing accuracy. Since the sensing information of different RSUs adopts different coordinate system, spatial registration for tracking-level or decision-level fusion is to adjust the sensing information of different RSUs to the same coordinate system, further to make up the sensing error and improve the sensing accuracy. 
	
	The development of spatial registration algorithms mainly goes through four stages. 
	
	The first stage is the spatial registration algorithms based on stereo projection. Both least square (LS) \cite{[LS]} and generalized least squares (GLS) \cite{[GLS]} are based on stereo projection. These algorithms project radar sensing information onto two-dimensional plane through conformal mapping, which is simple in principle and easy to implement. However, these algorithms do not consider the influence of errors caused by random noise on spatial registration. 
	
	The second stage is the maximum likelihood registration (MLR). MLR achieves the joint estimation of target state and system error by making the likelihood function reach the maximum and iterating recursively. However, MLR is an off-line spatial error registration algorithm, which is less real-time and less flexible \cite{[MLR_1],[MLR_2]}.
	
	The third stage is the spatial registration algorithms based on filtering algorithms. Typical filtering algorithms include the Kalman filter algorithm \cite{[KF]}, extended Kalman filter algorithm \cite{[EKF_UKF]}, unscented Kalman filter algorithm \cite{[EKF_UKF]}, etc. 
	These algorithms for online estimation add system error into the target state estimation by vector dimension extension of state vector. The system error and target state participate in filtering estimation together, so as to realize the joint estimation of target state and system error. However, these algorithms have the problem of high computational complexity.
	 
	With the rapid development of neural network, large-scale antenna and other technologies, spatial registration algorithms have entered the fourth stage, and some novel spatial registration algorithms gradually attracted the attention of researchers. 
	An adaptive error registration method based on neural networks was proposed in \cite{[NN_1]}. In the case of unknown prior errors, the algorithm learns the difference between unregistered target sensing results, without the need for complex registration system modeling. 
	J. Ma {\it et al.} proposed a new transformation estimation algorithm using the estimator and apply it to non-rigid registration for building robust sparse and dense mapping relations \cite{[2015]}.
	The generalized likelihood ratio test (GLRT) and Bayesian likelihood ratio test detectors are developed in \cite{[2016]} to achieve spatial registration.
	To boost registration algorithms, Parkison {\it et al.} developed a novel class of regularization models in the Reproducing Kernel Hilbert Space (RKHS) that ensures mapping relations are also consistent in an abstract vector space of functions such as intensity surface \cite{[2019_1]}.
	To solve the problems such as tracking registration failure and slow speed in the complex environment, a random fern classifier is applied to the process of the markerless augmented reality tracking registration in \cite{[2019_2]}. This method has better real-time performance than the traditional wide baseline matching algorithms.
	In order to reconstruct three-dimensional model, a registration process in three-dimensional space is required. For this problem, a three-dimensional registration method using the particle swarm optimization (PSO) is proposed in \cite{[2021]}. 
	The adjustable beam enabled spatial registration algorithm (AB-SRA) proposed in this paper realizes spatial registration by adjusting the beam width, which also benefits from the development of large-scale antenna technology.

	In recent years, cooperative sensing based on signal fusion has attracted more and more attention. In a cooperative radar sensing network with signal-level fusion, RSU directly combines multiple RSUs' echo signals to improve the sensing capability. In signal-level fusion, the Signal to Interference plus Noise Ratio (SINR) at the fusion center is improved compared with single radar sensing, so as to get better sensing capability. Compared with tracking-level and decision-level fusion, signal-level fusion does not require multiple extraction of information and can highly retain the effective sensing information in the original signal \cite{[signal_fusion_2020]}. 
	
	Since the sensing area of each RSU is different, the echo signal of different RSUs have spatial mismatch. Therefore, besides unifying coordinate systems, the aim of spatial registration in signal-level fusion is to make the sensing area of each RSU coincide, further to improve the target echo signal power and improve the target detection probability. So far, most of the research on signal fusion is achieved under the assumption that spatial registration had already been completed \cite{[no_signal_fusion]}. There are few researches on spatial registration algorithms in signal-level fusion. Therefore, it makes sense to propose an applicable solution for future engineering application. In order to realize the signal level fusion of the sensing information obtained from different RSUs, it is necessary to make the sensing regions of different RSUs highly consistent. 
	{
	Therefore, the main challenge of spatial registration for data-level fusion is spatial unification, which means to adjust the sensing information of different RSUs to achieve the unification of spatial coordinate system, further to make up the sensing error and improve the sensing accuracy. Compared with data-level fusion, the main challenge of spatial registration for signal-level fusion is spatial synchronization, which means to make the sensing area of each RSU coincide to achieve the synchronization of sensing area, further improving signal fusion, enhancing echo signals, and improving sensing performance.

	To overcome this challenge, Hong {\it{et al.}} proposed a spatial registration algorithm based on rasterizing the cooperative region \cite{[signal_fusion_space_reg]}. A relaxation parameter is put forward in the algorithm to balance the maximum SNR loss and the computation cost. But the relaxation parameter needs to be adjusted according to the radar distribution.
	Recognize this fact, this paper proposes AB-SRA that does not need to adjust the relaxation parameters. AB-SRA achieves spatial registration by adjusting the width of sensing beam to make the sensing area of each RSU coincide.}
	However, the increase of beam width will reduce the angle resolution. To mitigate the impact of reducing the angular resolution, we adjust the beam width while keeping the beam width as small as possible. To adjust the beamwidth flexibly, a constrained least squares problem is formulated to approach an ideal beampattern in \cite{[3DB_BF_1]}. Stoica {\it et al.} presented another optimization problem with the aim of producing a beam pattern with an expected 3dB main beam width \cite{[3DB_BF_2]}. In three-dimensional (3D) beam generation, the computational complexity of the above two algorithms is relatively high, a beamwidth adjustable beamforming algorithm (BABA) is proposed in this paper.
	
	Contributions of this paper are summarized as follows.
	
	1. We put forward a system model of RSU-CRSN with signal-level fusion. Different from tracking-level fusion and decision-level fusion, RSU directly combines multiple RSUs' echo signals for radar signal processing, which can improve the SINR of echo signals and thus improve the sensing capability.
	
	2. We present an adjustable beam enabled spatial registration algorithm, named AB-SRA, that makes the sensing area of each RSU coincide by adjusting the sensing beam width of the RSU. The algorithm only needs to adjust the width and direction of the sensing beam according to the relative position information of RSUs and targets. AB-SRA has a low complexity and does not need to adjust the parameters according to the deployment changes of RSUs. Moreover, we further provide the performance comparison and analysis between AB-SRA and the typical spatial registration algorithm proposed in \cite{[signal_fusion_space_reg]}. Compared with the spatial registration algorithm proposed in \cite{[signal_fusion_space_reg]}, AB-SRA is more suitable for scenarios where the requirements of the sensing angular resolution is not high but that of the sensing time is high.

	3. We propose a beamforming algorithm which can flexibly adjust the beam width. Compared with the traditional beamforming algorithm, the proposed algorithm optimizes the beam direction and width in two steps, and it does not need to transform the problem into convex optimization.
	
	The remaining parts of this paper are organized as follows.
	Section \ref{sec:system-model} describes the system model of RSU-CRSN. Section \ref{sec:registration-algorithm} elaborates AB-SRA. The design of BABA is introduced in section \ref{sec:beamforming-algorithm}. In section \ref{sec:ROC}, the sensing capability of RSU-CRSN is derived in detail. The performance of BABA and AB-SRA is simulated in section \ref{sec:Radar-Performance}. Section \ref{sec:Conclusion} concludes the paper. 
	
	The symbols used in this paper are described as follows. Vectors and matrices are denoted by boldface small and capital letters; the transpose, complex conjugate, Hermitian, inverse, and pseudo-inverse of the matrix ${\bf{A}}$ are denoted by ${{\bf{A}}^T}$, ${{\bf{A}}^*}$, ${{\bf{A}}^H}$, ${{\bf{A}}^{ - 1}}$ and ${{\bf{A}}^\dag}$, respectively; ${\rm{diag}}(\bf{x})$ is the operation that generates a diagonal matrix with the diagonal elements to be the elements of $\bf x$; $ \otimes $ is the Kronecker product operator; $ \odot $ is the Hamdard product operator.
	
	Key parameters and abbreviations in this
	paper are given in Table \ref{label:abbreviations}.
	
	\begin{table}[t]
		\caption{Key Parameters and Abbreviations}
		\label{label:abbreviations}
		\begin{tabular}{l|l|l|l}
			\hline \hline
			Abbreviation & Description & Abbreviation & Description  \\ \hline
			ITS &  Intelligent transportation system & RSU & Roadside unit\\ \hline
			RSU-CRSN & Multiple RSUs cooperative radar sensing network & AB-SRA &  Adjustable beam enabled spatial registration algorithm \\ \hline
			BABA & Beamwidth adjustable beamforming algorithm & CSA & cooperative sensing area \\ \hline
			DFC & Data fusion center & PFA & Phase reference antenna \\ \hline
			LCMV & Linear constrained minimum variance & ROC & Receiver operating characteristic \\ \hline
			3D & Three-dimensional & HBF & hybrid analogue and digital beamforming  \\ \hline
			DBF & Digital beamforming & DAC & Digital-to-Analogue Conversion\\ \hline
			RF & Radio frequency & PA & Phase amplification  \\ \hline
			${{\it{f}}_{\rm{sb}}}({\it{r}},\theta ,\phi )$ & Function of the sensing beam & ${{\it{f}}_{\rm{csa}}}({\it{r}},\theta ,\phi )$ & Function of CSA\\ \hline
			${{\it{f}}_{\it{u}}}({\it{r}},\theta ,\phi )$ & Function of the sensing unit & $h$ & Height of the RSU antenna array \\ \hline
			$r_a$ & Major semiaxis of sensing unit & $r_b$ & Minor semiaxis of sensing unit \\ \hline
			$R_t$ & Distance between the target and the RSU	antenna array & $e_b$ & Performance of BABA \\ \hline
			$p_d$ & Detection probability & $p_f$ & False alarm
			probability \\ \hline  
			$P_{\rm{dfc}}$ & Matching degree of CSA of each RSU \\ \hline
		\end{tabular}
	\end{table}

	\section{Problem Formulation}\label{sec:system-model}
	
	In this section, we first describe the cooperative sensing system model and signal model for cooperative sensing. We then formulate the spatial registration problem of cooperative sensing.
	
	\subsection{Cooperative Sensing System Model}\label{sec:sys}
	\begin{figure}[ht]
		\includegraphics[scale=0.4]{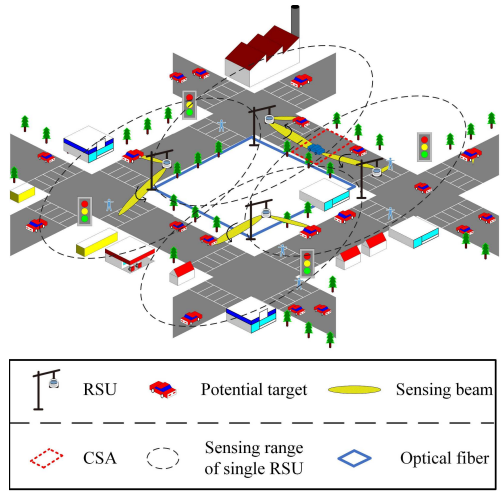}
		\centering
		\caption{RSU-CRSN system.}
		\label{fig:RSU-CRSN}
	\end{figure}
	
	As Fig. \ref{fig:RSU-CRSN} shows, the antenna array mounted on RSU can form multiple radar sensing beams to detect targets on the ground. The radar sensing range of single RSU is denoted by the gray dotted line. To detect the target within the sensing range, RSU needs to generate sensing beam pointing to it. Since SNR at the targets on the edge of the sensing range is low, the sensing performance is poor. So multiple RSUs can cooperate to sense through signal-level cooperation to improve the sensing performance. As Fig. \ref{fig:RSU-CRSN} shows, two RSUs are cooperatively sensing the blue vehicle, the red dotted line denotes the cooperative sensing area (CSA). 
	{It should be noted that we focus on urban traffic environments, where vehicle speeds are mostly in the range of $20 \sim 40$ km/h.}
	
	When RSU receives a cooperative sensing request, it will first send the location of the center of CSA to other RSUs. Then, each RSU forms a sensing beam pointing to CSA, and transmit echo signals to the data fusion center (DFC). Finally, DFC gets sensing information from the fusion echo signals.
	Signal-level cooperative sensing requires highly precise synchronization between RSUs. As Fig. \ref{fig:RSU-CRSN} shows, RSUs are connected by optical fibers for two purposes: 1) to reduce the transmission delay of sensing signals; 2) to make the sensing signals from each RSU reach DFC as synchronized as possible.
	Synchronization of cooperative sensing among RSUs essentially requires that the time when the sensing signals of each RSU reach the fusion center is synchronized. To satisfy this purpose, it is not only required that each RSU transmits sensing signals simultaneously, but also requires that the transmission delay of each sensing signal from RSU to DFC is as low as possible. To satisfy the first requirement, it is necessary to ensure the clock synchronization between RSUs. For bi-static and multi-static radars, clock synchronization is typically realized via wired connections or locking to the GPS signals \cite{[AJZ_1],[AJZ_2]}. To solve the clock synchronization problem between RSUs, serial digital interface (SDI) high-precision time synchronization equipment based on BeiDou or GPS can be used, and synchronization signals can be transmitted wirelessly to complete synchronization coordination between multiple radars \cite{[Syn_1],[Syn_2]}. To satisfy the second requirement, it is necessary to use a large bandwidth and low delay communication transmission. Fiber optical communication is a viable option. Optical fiber connection has been applied in signal-level fusion multi-radar networking \cite{[SL_1]}. According to \cite{[SL_1]}, the radar networking of signal-level fusion requires the transmission delay between
	RSUs to be at the $ms$ level and the communication rate to be at the $Gb/s$ level. RSUs can be connected through gigabit optical fibers to meet the above requirements.

	\begin{figure}[t]
		\includegraphics[scale=0.3]{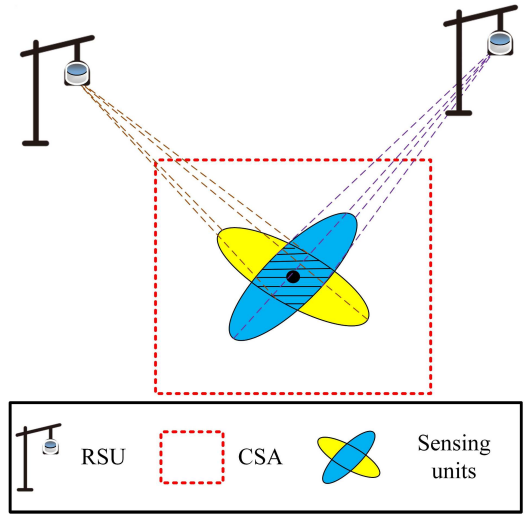}
		\centering
		\caption{Sensing units before spatial registration.}
		\label{fig:space_1}
	\end{figure}
		
	\begin{figure}[b]
		\includegraphics[scale=0.3]{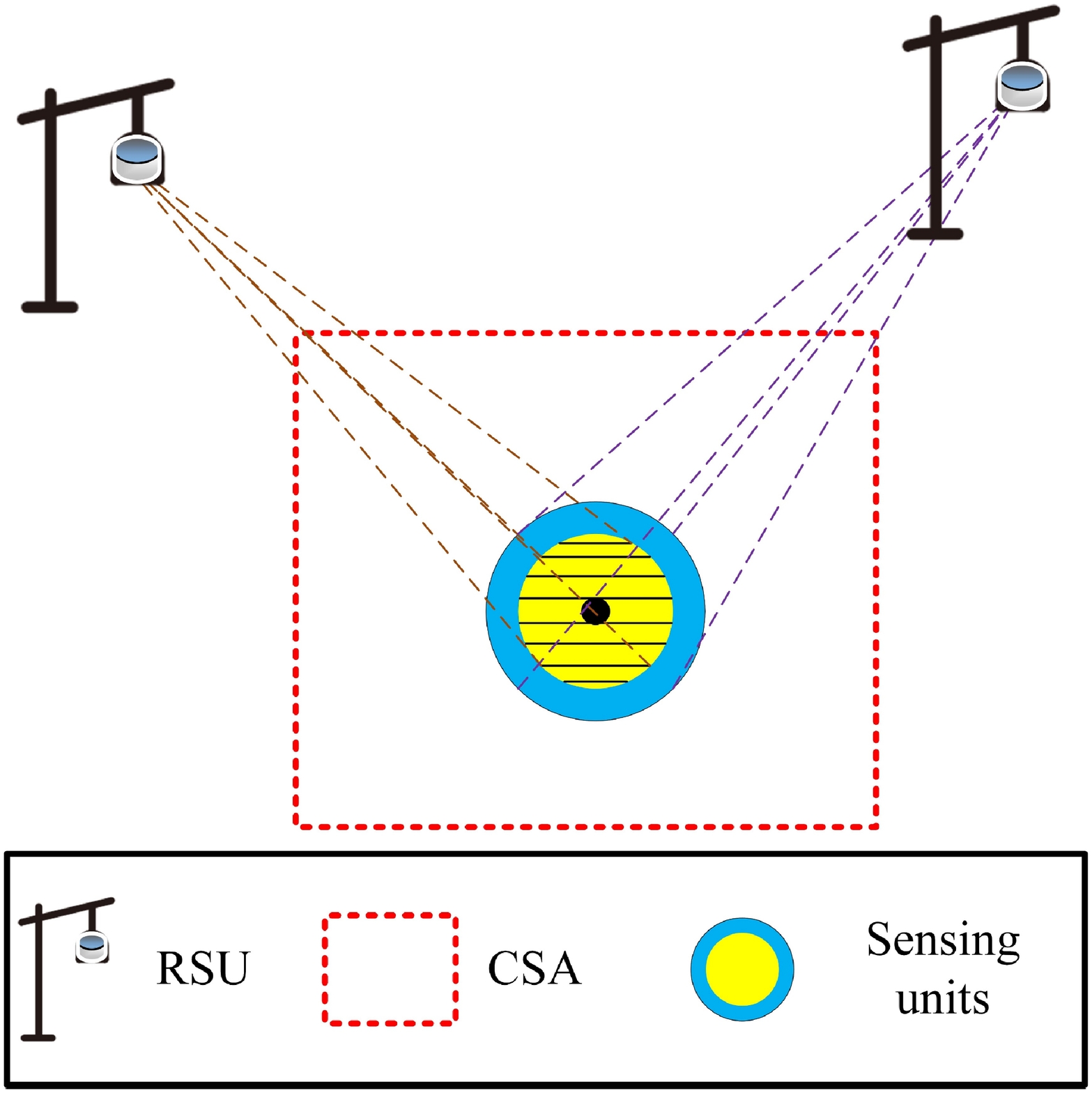}
		\centering
		\caption{Spatial registration of a single RSU.}
		\label{fig:space_2}
	\end{figure}
	
	\begin{figure}[b]
		\includegraphics[scale=0.3]{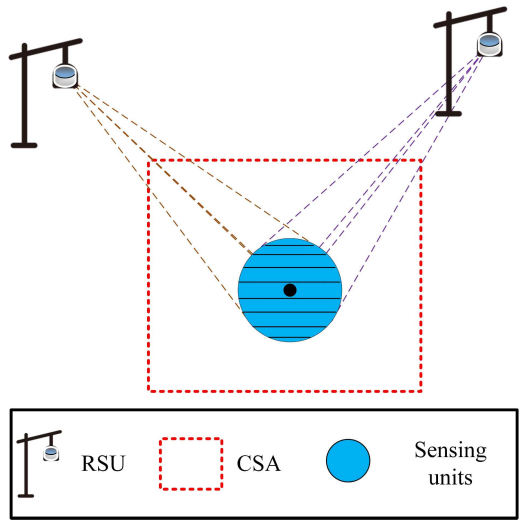}
		\centering
		\caption{Spatial registration of all cooperative RSUs.}
		\label{fig:space_3}
	\end{figure}

	\subsection{Signal Model for Cooperative Sensing}\label{sec:CSM}
	
	For convenience of description, we define the coverage of single sensing beam on CSA as the sensing unit for short. Different RSUs will generate different sensing units, the overlapping area of sensing units is the area where the target echo signal is superimposed and enhanced. 
	Assuming that RSU realizes signal-level fusion by means of simple incoherent accumulation \cite{[incoherent]}, the average echo signal obtained by RSU is $P_{\rm{dfc}} \cdot \sum_{i} S_i$, where $S_i$ is the echo signal of each RSU, $P_{\rm{dfc}} \in (0,1)$ represents the matching degree of the sensing units of different RSUs. 
	Then, when there is a target, the echo signal of RSU after signal fusion can be expressed as
	\begin{equation}
	y = P_{\rm{dfc}} \cdot \sum_{i} S_i + n + I,
	\label{equ:received-echo-signal}
	\end{equation}
	where $n$ and $I$ denote noise and ground clutter, respectively.
	The more closely the sensing units of different RSUs overlap, the larger the signal enhancement area will be, and the larger $P_{\rm{dfc}}$ will be. In extreme cases, $P_{\rm{dfc}} = 0$ means no registration at all, $P_{\rm{dfc}} = 1$ means perfect registration.
	
	\subsection{Spatial Registration Problem of Cooperative Sensing}\label{sec:SP-CSM}
	
	In order to get better sensing performance, higher SINR is needed. Then the spatial registration problem of cooperative sensing can be expressed as
	\begin{equation}
	\max_{{P_{\rm{dfc}}}} {\frac{P_{\rm{dfc}} \cdot \sum_{i} S_i}{n+I}}.
	\end{equation}
	
	If the sensing unit of each RSU is coincident, multiple RSUs get echo signals reflected from the same sensing area, which is the goal of spatial registration in signal-level fusion. Since the locations of RSUs are different, sensing units of different RSUs are not coincident in terms of the direction and size, the received sensing SINR is significantly reduced. 
	
	As Fig. \ref{fig:space_1} shows, two sensing units with a certain angle offset are not completely coincident. To get higher ${P_{\rm{dfc}}}$ after signal-level fusion, we propose AB-SRA to register the sensing units.
	
	\section{AB-SRA}\label{sec:registration-algorithm}

	AB-SRA is proposed to realize spatial registration, makes the sensing units of each RSU coincide by adjusting the width of the RSUs' sensing beams.
	It is implemented in two stages. In the first stage, RSU makes the sensing unit be an approximate circularity by adjusting the width of sensing beam. As Fig. \ref{fig:space_2} shows, if each sensing unit has a circular shape, they will match with each other to the greatest extent.
	
	However, sensing units are not completely matched yet because the size of sensing units is different. In the second stage, RSU make the size of sensing units be the same by adjusting the width of sensing beams. So far, sensing units of different RSUs are completely matched, as Fig. \ref{fig:space_3} shows.
	
	\subsection{Details of AB-SRA}\label{sec:AB-SRA_1}
    
    This subsection will introduce four main steps of AB-SRA: 1) coordinate system unification, 2) CSA identification, 3) sensing unit simplification, and 4) sensing beam adjustment.

	\subsubsection{Coordinate System Unification}
	
	Coordinate system unification is the prerequisite for the unified positioning of the sensing unit and the determination of sensing beam direction. In the RSU-CRSN system, different RSUs adopt different coordinate systems. It is necessary to uniform coordinate systems before data fusion. There are some widely used coordinate systems, such as cartesian coordinate system, spherical coordinate system, polar coordinate system and earth-centered earth-fixed coordinate system \cite{[ECEF_coordinate_system]}. Some conversion formulas between different coordinate systems are presented in \cite{[coordinate_system_change]}.
	
	RSU mainly adopts two coordinate systems: cartesian coordinate system and spherical/polar coordinates system. In the cartesian coordinate system, the location of RSU is defined as the origin. The horizontal plane of RSU is defined as the $xoy$ plane. The $x$ axis and the platform axis are pointed forward in the same direction and perpendicular to the $y$ axis. The positive direction of the $z$ axis is perpendicular to the $xoy$ plane. The position of the target is expressed as $(x,y,z)$. In the spherical/polar coordinates system, the position of the target is expressed as $(r,\theta ,\phi )$, where $r$, $\theta$ and $\phi$ denote the radial distance, pitch angle and azimuth angle measured by RSU to the target, respectively. The coordinate transformation formula of the above two coordinate systems can be expressed as 
	\begin{equation}\label{equ:coordinate_transformation}
	\left[ {\begin{array}{*{20}{c}}
		x\\
		y\\
		z
		\end{array}} \right] = \left[ {\begin{array}{*{20}{c}}
		{r\sin \theta \cos \phi }\\
		{r\sin \theta \sin \phi }\\
		{r\cos \theta }
		\end{array}} \right].
	\end{equation}
	
	\subsubsection{CSA Identification}
	
    CSA is a target-centered square area with side length ${r_d}$. CSA Identification after unifying coordinate systems, each RSU participating in cooperative sensing can identify CSA in the same coordinate system. Assuming that there are $k$ RSUs participated in cooperative sensing, denoted by ${{\it{P}}_1},{{\it{P}}_2},...,{{\it{P}}_k}$. The set of the locations of targets relative to all RSUs can be expressed as ${{\it{S}}_e}{\rm{ = \{ (}}{{\it{R}}_i}{\rm{,}}{\theta _i}{\rm{,}}{\varphi _i}{\rm{) | {\it i} = 1,2,}}...{\it{,k\} }}$, where ${\rm{(}}{{\it{R}}_i}{\rm{,}}{\theta _i}{\rm{,}}{\varphi _i}{\rm{)}}$ denotes the position of target relative to ${\it P_{\it i}}$.
	
	\subsubsection{Sensing Unit Simplification} 
	
\begin{figure}[ht]
		\includegraphics[scale=0.3]{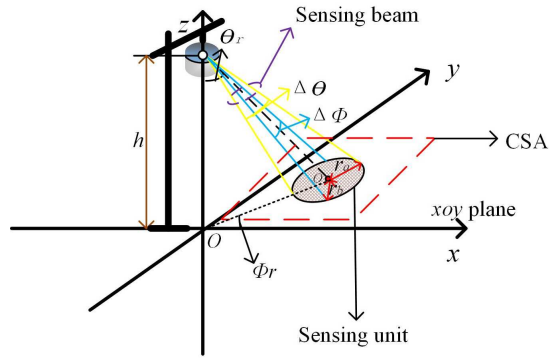}
		\centering
		\caption{Single sensing beam coverage.}
		\label{fig:single_beam_area}
	\end{figure}

	The shape of the sensing unit formed by the 3D sensing beam and CSA is irregular. In order to facilitate theoretical derivation without losing the truth, we simplified the sensing unit. It mainly includes two parts: CSA simplification and sensing beam simplification. As Fig. \ref{fig:single_beam_area} shows, the sensing unit is generated by the intersection of CSA and 3D sensing beam. In the spherical/polar coordinates system, the function of the sensing beam and CSA can be respectively expressed as ${{\it{f}}_{\rm{sb}}}({\it{r}},\theta ,\phi )$ and ${{\it{f}}_{\rm{csa}}}({\it{r}},\theta ,\phi )$. So that, the sensing unit can be deduced as
    \begin{equation}
    \left\{
    \begin{aligned}
    &{{{\it{f}}_{\rm{sb}}}({\it{r}},\theta ,\phi )}\\
    &{{{\it{f}}_{\rm{csa}}}({\it{r}},\theta ,\phi )}
    \end{aligned}. \right.
    \label{equ:DU_1}
    \end{equation}
	In order to solve \eqref{equ:DU_1}, two-step model simplification is adopted.
	
	$\bf{CSA~simplification:}$
	
	 Considering that a steep road may cause vehicle field of vision blind area, there is rarely road with big radian in a real traffic environment. Moreover, the size of CSA is generally small, so that CSA can be approximated to a plane without radian bending. Then ${{\it{f}}_{\rm{csa}}}({\it{r}},\theta ,\phi )$ in the Cartesian coordinate system can be expressed as
	 \begin{equation}\label{equ:plane_add}
	 A x  + B y  + C z  + E{\rm{ = 0}},
	 \end{equation}
	 where $A$, $B$, $C$, and $E$ are the undetermined parameters.
	 According to the coordinate transformation formula \ref{equ:coordinate_transformation}, ${{\it{f}}_{\rm{csa}}}({\it{r}},\theta ,\phi )$ in the spherical/polar coordinates system can be expressed as 
	 \begin{equation}\label{equ:plane}
	 {\it r}(A \sin \theta \cos \phi  + B \sin \theta \sin \phi  + C\cos \theta ) + E{\rm{ = 0}},
	 \end{equation}

	$\bf{Sensing~beam~simplification:}$ 
	
	As Fig. \ref{fig:single_beam_area} shows, sensing beam emitted from the RSU antenna array at $(0,0,\it h)$ is conical. So that, ${{\it{f}}_{\rm{sb}}}({\it{r}},\theta ,\phi )$ can be expressed as
	\begin{equation}\label{equ:3D_beam}
	\begin{array}{c}
	{{\rm{({\it r} \cos(}}{\theta _{rot}}{\rm{) - {\it h})}}^2}{\rm{ - (}}\frac{{{{(r{\sin}({\theta _{rot}}){ \cos}\phi )}^2}}}{{{{\tan }^2}(\frac{{\Delta \theta }}{2})}}{\rm{ + }}\frac{{{{(r{\sin}({\theta _{rot}}){\sin}\phi )}^2}}}{{{{\tan }^2}(\frac{{\Delta \phi }}{2})}}{\rm{) = 0}},\\ \\
	{\theta _{rot}} = \theta {\rm{ - }}{\theta _r}{\rm{ - }}\frac{\pi }{2},
	\end{array}
	\end{equation}
	where $({\theta _r},{\phi _r})$ is the direction of sensing beam, $(\Delta \theta ,\Delta \phi )$ denotes the width of sensing beam.
	
	In order to simplify \eqref{equ:3D_beam}, some approximations are applied in this paper. For small size targets, the sensing beam can be regarded as being projected parallel to the sensing unit. Therefore, the sensing beam can be modeled by a cylinder. Then, the sensing unit can be simplified as follows
	\begin{equation}\label{equ:DU_2}
	\begin{array}{c}
	{\rm{(}}\frac{{{{(r{\rm sin}({\theta _{rot}}){\cos}\phi )}^2}}}{{{{\tan }^2}(\frac{{\Delta \theta }}{2})}}{\rm{ + }}\frac{{{{(r{\rm sin}({\theta _{rot}}){\rm sin}\phi )}^2}}}{{{{\tan }^2}(\frac{{\Delta \phi }}{2})}}{\rm{) = }}{{\it{R}}^2_{\it{t}}},\\ \\
	{\theta _{rot}} = \theta {\rm{ - }}{\theta _r}{\rm{ - }}\frac{\pi }{2},\\ \\
	r(A{\rm sin}\theta \cos \phi  + B\sin \theta \sin \phi  + C\cos \theta ) + {\it{E = 0}},
	\end{array}
	\end{equation}
	where ${{\it{R}}_{\it{t}}}$ is the distance between the target and the RSU antenna array.
	
	\subsubsection{Sensing Beam Adjustment}
	
	Sensing beam adjustment is the main step of AB-SRA. The sensing units of each RSU are coincided via sensing beam adjustment.
	Sensing beam adjustment can be divided into two sub-steps. The first sub-step is to make sensing units be approximately circular. As Fig. \ref{fig:single_beam_area} shows, the sensing unit is an ellipse with major semiaxis ${{\it{r}}_a}$ and minor semiaxis ${{\it{r}}_b}$. If ${{\it{r}}_a} = {{\it{r}}_b}$, the sensing unit will be a circularity. The second sub-step is to make all sensing units be in the same size.
	
	\subsection{Angular Resolution after AB-SRA}
	
	\begin{figure}[ht]
		\includegraphics[scale=0.25]{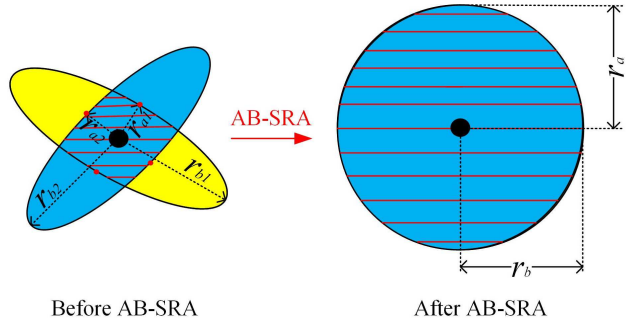}
		\centering
		\caption{Angular resolution before and after AB-SRA.}
		\label{fig:Angular_Resolution}
	\end{figure}
	
	It could be noted that AB-SRA only needs to adjust the width and direction of the sensing beam, carry out simple geometric operations, and the algorithm complexity is relatively low. 
	However, although AB-SRA achieves spatial registration at the minimum beamwidth, increasing the beamwidth will sacrifice the angular resolution of the sensor to some extent.
	This subsection will briefly analyze the effect of adjusting the beam on angular resolution.
	
	For the convenience of description, this subsection takes the cooperative sensing of two RSUs as an example. According to \eqref{equ:3D_beam_1}, the angular resolution $\Delta \theta$ and $\Delta \phi$ can be expressed by the size of sensing unit $r_a$ and $r_b$.
	Fig. \ref{fig:Angular_Resolution} shows the angular resolution before and after AB-SRA. Assuming that the size of sensing unit of two RSUs is $r_{a1}$, $r_{b1}$, $r_{a2}$ and $r_{b2}$. Without loss of generality, assuming that $r_{b1} \ge r_{a2}$ and $r_{b2} \ge r_{a1}$.
	
	According to section \ref{sec:AB-SRA_1}, the size of sensing unit after AB-SRA can be expressed as
	\begin{equation}\label{equ:AR_1}
	\begin{aligned}
	r_a &= max\{ r_{a1}, r_{a2}, r_{b1}, r_{b2}  \} \\
	r_b &= r_a
	\end{aligned}.
	\end{equation}
	
	The size of sensing unit before AB-SRA can be expressed differently according to different detection results of RSUs.
	Case 1: 
	If both RSUs detect the target, it can be judged that the target is in the overlapping area of the sensing units of the two RSUs, then the size of sensing unit before AB-SRA can be expressed as
	\begin{equation}\label{equ:AR_2}
	\begin{aligned}
	r_a &= min\{ r_{a1}, r_{b2}  \} \\
	r_b &= min\{ r_{a2}, r_{b1}  \} \\
	\end{aligned}.
	\end{equation}
	Case 2:
	If RSU 1 detects the target, but RSU 2 does not, the size of sensing unit before AB-SRA can be expressed as
	\begin{equation}\label{equ:AR_3}
	\begin{aligned}
	r_a &= max\{r_{b1}, r_{a2}\} - min\{r_{b1}, r_{a2}\} \\
	r_b &= max\{r_{b2}, r_{a1}\} - min\{r_{b2}, r_{a1}\} \\
	\end{aligned}.
	\end{equation}
	Case 3:
	If RSU 2 detects the target, but RSU 1 does not, the size of sensing unit before AB-SRA can be expressed as
	\begin{equation}\label{equ:AR_4}
	\begin{aligned}
	r_a &= max\{r_{b2}, r_{a1}\} - min\{r_{b2}, r_{a1}\} \\
	r_b &= max\{r_{b1}, r_{a2}\} - min\{r_{b1}, r_{a2}\} \\
	\end{aligned}.
	\end{equation}
	
	Compared \eqref{equ:AR_2}, \eqref{equ:AR_3}, \eqref{equ:AR_4} and \eqref{equ:AR_1}, it could be noted that AB-SRA will sacrifice the angular resolution of the sensing beam to some extent. 
	However, it is acceptable to obtain a higher detection probability of target at the expense of certain angular resolution for the scenario of collaborative sensing with small echo signal and low detection probability of edge target.    
	In real applications, we need to balance the performance of angular resolution and spatial registration to choose an appropriate beamwidth.  
	
	\subsection{Sensing Beam Optimization in Specific Scenarios} \label{subsec:registration-algorithm_Specific_Scenarios}
	
	Different CSAs have different ${{\it{f}}_{\rm{csa}}}({\it{r}},\theta ,\phi )$. Several common CSAs will be discussed in this subsection. Model 1 represents the case where CSA is just parallel to the $xoy$ plane. Model 2 represents the case where CSA is not parallel to the $xoy$ plane. \eqref{equ:DU_1} will be solved to obtain the limitation of sensing beam's width.
	
	\subsubsection{Model 1} 
	
	As Fig. \ref{fig:single_beam_area} shows, assuming that CSA is just parallel to the $xoy$ plane that is perpendicular to RSU, then ${{\it{f}}_{\rm{csa}}}({\it{r}},\theta ,\phi )$ can be expressed as
	\begin{equation}\label{equ:DU_3}
	{\it{r}}\cos \theta  = {\it{h}} - {\it{R}_t}\cos {\theta _r},
	\end{equation}
	where ${\it{h}}$ is the height of the RSU antenna array. Further, the sensing unit can be expressed as
	\begin{equation}\label{equ:DU_4}
	\begin{array}{c}
	{\rm{(}}\frac{{{{(r{\sin}({\theta _{rot}}){\cos}\phi )}^2}}}{{{{\tan }^2}(\frac{{\Delta \theta }}{2})}}{\rm{ + }}\frac{{{{(r{\sin}({\theta _{rot}}){\sin}\phi )}^2}}}{{{{\tan }^2}(\frac{{\Delta \phi }}{2})}}{\rm{) = }}{{\it{R}}^2_{\it{t}}},\\ \\
	{\theta _{rot}} = \theta {\rm{ - }}{\theta _r}{\rm{ - }}\frac{\pi }{2},\\ \\
	{\it{r}}\cos \theta  = {\it{h}} - {\it{R}_t}\cos {\theta _r}.
	\end{array}
	\end{equation}

	After solving \eqref{equ:DU_4}, the sensing unit marked as ${{\it{f}}_u}(r,\theta ,\phi )$ can be derived as
	\begin{equation}\label{equ:DU_5}
	\begin{array}{c}
	{\rm{(}}\frac{{{{(({\it{h}} - {\it{R}_t}\cos {\theta _r}) \cdot {\sin}(\theta  - {\theta _r} - \frac{\pi }{2}){\cos}\phi )}^2}}}{{{{{\rm{(}}{{\it{R}}_{\it{t}}}{\cos}\theta  \cdot \tan (\frac{{\Delta \theta }}{2}))}^2}}}\\ \\
	{\rm{ + }}\frac{{{{(({\it{h}} - {\it{R}_t}\cos {\theta _r}) \cdot {\sin}(\theta  - {\theta _r} - \frac{\pi }{2}){\sin}\phi )}^2}}}{{{{{\rm{(}}{{\it{R}}_{\it{t}}}{\cos}\theta  \cdot \tan (\frac{{\Delta \phi }}{2}))}^2}}}{\rm{) = }}1.
	\end{array}
	\end{equation}
	Obviously, ${{\it{f}}_u}(r,\theta ,\phi )$ is an equation for an ellipse. The major semiaxis ${{\it{r}}_a}$ and minor semiaxis ${{\it{r}}_b}$ can be derived as follows
	\begin{equation}\label{equ:ra_rb_1}
	\begin{array}{c}
	{{\it{r}}_a} = \frac{{{\it{R}_t}\tan (\frac{{\Delta \theta }}{2})}}{{\cos {\theta _r}}},\\ \\
	{{\it{r}}_b} = {\it{R}_t}\tan (\frac{{\Delta \phi }}{2}).
	\end{array}
	\end{equation}
	
	Based on \eqref{equ:ra_rb_1}, the limitation of sensing beam can be obtained by solving the following equation
	\begin{equation}\label{equ:ra_rb_2}
	{{\it{r}}_a} = \frac{{{\it{R}_t}\tan (\frac{{\Delta \theta }}{2})}}{{\cos {\theta _r}}}{\rm{ = }}{{\it{r}}_b} = {\it{R}_t}\tan (\frac{{\Delta \phi }}{2}).
	\end{equation}
	The solution of \eqref{equ:ra_rb_2} is shown as
	\begin{equation}\label{equ:limitation_1}
	\Delta \theta {\rm{ = }}2\arctan ({\cos}{\theta _r} \cdot {\tan}(\frac{{\Delta \phi }}{2})).
	\end{equation}
	After adjusting the width of sensing beams according to \eqref{equ:limitation_1}, ${{\it{S}}_{\it{e}}}$ can be modified as ${\rm{ \{ (}}{{\it{R}}_i}{\rm{,}}{\theta _i}{\rm{,}}{\phi _i}{\rm{,}}\Delta {\theta _i},\Delta {\phi _i}{\rm{) | {\it{i}} = 1,2,}}...{\it{,k\} }}$, where $(\Delta {\theta _i},\Delta {\phi _i})$ is the width of sensing beam generated by ${\it P}_{\it i}$.

	Position information set ${{\it{S}}_{\it{e}}}$ is sorted by the detection distance ${{\it{R}}_{\it{i}}}$. The middle of ${{\it{S}}_{\it{e}}}$ is chosen as the reference sensing beam, which is marked as ${\rm{(}}{{\it{R}}_{\it{p}}}{\rm{,}}{\theta _p}{\rm{,}}{\phi _p}{\rm{,}}\Delta {\theta _p},\Delta {\phi _p}{\rm{)}}$. If adjust the width of sensing beam to meet \eqref{equ:limitation_2}, the size of sensing units generated by different RSUs will be the same.
	\begin{equation}\label{equ:limitation_2}
	{{\it {R}}_i}\tan (\frac{{\Delta {\phi _{\it{i}}}}}{2}) = {{\it {R}}_p}\tan (\frac{{\Delta {\phi _p}}}{2}).
	\end{equation}
	
	Sensing beam optimization is achieved by adjusting the beamwidth to satisfy \eqref{equ:limitation_1} and \eqref{equ:limitation_2}.
	
	\subsubsection{Model 2} 
	
	Assuming that CSA and the $xoy$ plane present a certain angle ${\theta _{\it{t}}}$, ${{\it{f}}_{\rm{csa}}}({\rm{r}},\theta ,\phi )$ can be expressed as
	\begin{equation}\label{equ:DU_6}
	{\it{r}}\cos (\theta  - {\theta _t}) = {\it{h}} - {{\it R}_t}\cos {\theta _r}.
	\end{equation}
	The sensing unit can be expressed as
	\begin{equation}\label{equ:DU_7}
	\begin{array}{c}
	{\rm{(}}\frac{{{{(r{\sin}({\theta _{rot}}){\cos}\phi )}^2}}}{{{{\tan }^2}(\frac{{\Delta \theta }}{2})}}{\rm{ + }}\frac{{{{(r{\sin}({\theta _{rot}}){\sin}\phi )}^2}}}{{{{\tan }^2}(\frac{{\Delta \phi }}{2})}}{\rm{) = }}{{\it{R}}^2_{\it{t}}},\\ \\
	{\theta _{rot}} = \theta {\rm{ - }}{\theta _r}{\rm{ - }}\frac{\pi }{2},\\ \\
	{\it{r}}\cos (\theta  - {\theta _t}) = {\it{h}} - {{\it R}_t}\cos {\theta _r}.
	\end{array}
	\end{equation}
	The solution of \eqref{equ:DU_7} is
	\begin{equation}\label{equ:DU_8}
	\begin{array}{c}
	{\rm{(}}\frac{{{{(({\it{h}} - {{\it R}_t}\cos {\theta _r}) \cdot {\sin}(\theta  - {\theta _r} - \frac{\pi }{2}){\cos}\phi )}^2}}}{{{{{\rm{(}}{{\it{R}}_{\it{t}}}{\cos}(\theta  - {\theta _t}) \cdot \tan (\frac{{\Delta \theta }}{2}))}^2}}}\\ \\
	{\rm{ + }}\frac{{{{(({\it{h}} - {{\it R}_t}\cos {\theta _r}) \cdot {\sin}(\theta  - {\theta _r} - \frac{\pi }{2}){\sin}\phi )}^2}}}{{{{{\rm{(}}{{\it{R}}_{\it{t}}}{\cos}(\theta  - {\theta _t}) \cdot \tan (\frac{{\Delta \phi }}{2}))}^2}}}{\rm{) = }}1.
	\end{array}
	\end{equation}
	
	Therefore, the major semiaxis ${{\it{r}}_a}$ and minor semiaxis ${{\it{r}}_b}$ can be derived as
	\begin{equation}\label{equ:ra_rb_3}
	\begin{array}{l}
	{{\it{r}}_a} = \frac{{{{\it R}_t}\tan (\frac{{\Delta \theta }}{2})}}{{\cos ({\theta _r} - {\theta _{\it{t}}})}},\\ \\
	{{\it{r}}_b} = {{\it R}_t}\tan (\frac{{\Delta \phi }}{2}).
	\end{array}
	\end{equation}
	The rest of the optimizations are the same as Model 1.
	
 	\section{Beamforming Algorithm}\label{sec:beamforming-algorithm}
	
	In this section, we first present the RSU antenna array and improved LCMV beamforming algorithm. We then propose BABA base on DBF and HBF, respectively.
	
	\subsection{Design of the RSU Antenna Array}\label{RSU:Antenna}
	In order to generate 3D downward sensing beam, the circular center antenna array is adopted in the RSU-CRSN system \cite{[ant_1]}. 
	
\begin{figure}[ht]
		\includegraphics[scale=0.22]{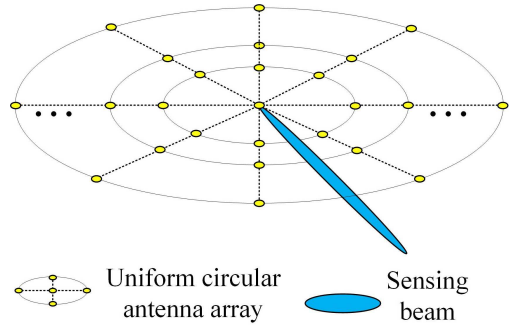}
		\centering
		\caption{RSU-CRSN antenna array.}
		\label{fig:ant_1}
	\end{figure}

\begin{figure}[ht]
		\includegraphics[scale=0.25]{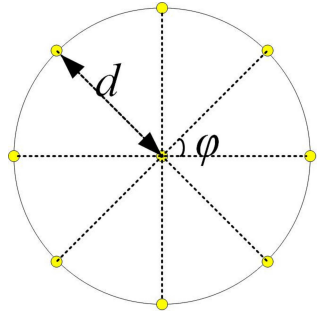}
		\centering
		\caption{Single layer of RSU-CRSN antenna array.}
		\label{fig:ant_3}
	\end{figure}

	As Fig. \ref{fig:ant_1} shows, there are several layers in the array. The layer with index $\it i$ is labeled as the $\it i$th layer. The center antenna is selected as the phase reference antenna (PFA). From the center to the periphery, the layers are labeled from $0$ to $p-1$. There are ${2^b}$ antenna elements in each layer, except the $\it 0$th layer that has only one antenna element. The distance between two adjacent layers is ${\it{d}}$. Details of antenna elements in a single layer are shown in Fig. \ref{fig:ant_3}.
	
	In order to avoid phase ambiguity, the distance between adjacent antenna elements should meet the following conditions
	\begin{equation}\label{equ:avoid_phase_ambiguity}
		\begin{array}{c}
			{\it{d}} \le {\rm{1/2}}\lambda, \\
			2{\it{d}}\sin \frac{\varphi }{2} \le 1/2\lambda,
		\end{array}
	\end{equation}
	where $\varphi {\rm{ = 2}}\pi {\rm{/(}}{{\rm{2}}^b}{\rm{)}}$ is the angle difference of adjacent elements within the same layer. Condition \eqref{equ:avoid_phase_ambiguity} indicates that both the distance between adjacent layers and the distance between adjacent elements within the same layer should be smaller than half of the wavelength.

	\subsection{Improved LCMV Beamforming Algorithm}\label{I-LCMV}
	
	\begin{figure}[ht]
		\includegraphics[scale=0.27]{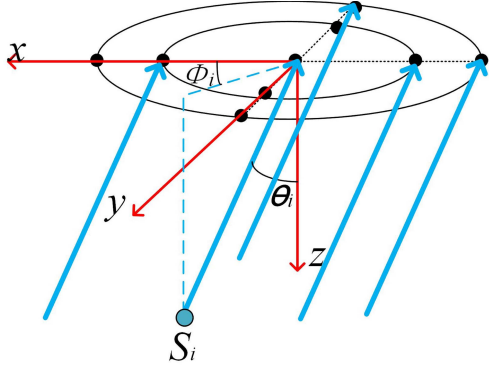}
		\centering
		\caption{Receiving signals of RSU-CRSN antenna.}
		\label{fig:ant_2}
	\end{figure}
	
	As Fig. \ref{fig:ant_2} shows, assuming that there are ${\it{l}}$ far field signals coming from different directions. The $\it{i}$th planar far field signal comes from the direction of $({\phi _i},{\theta _i})$. The phase difference between the ${\it{n}}$th antenna element of the ${\it{m}}$th layer and PFA caused by the ${\it{i}}$th planar far field signal can be expressed as
	\begin{equation}\label{equ:daoxiang}
	{{\it{a}}_{m,n}}({\phi _i},{\theta _i}) = exp( - j\frac{{2\pi }}{\lambda } {\bf{q}} _{m,n}^T{ {\bf{v}} _i}),
	\end{equation}
	where $\lambda$ is the signal wavelength, the polar mapping vector ${{\bf{v}} _i}$ can be described as
	\begin{equation}\label{equ:polar_mapping_vector}
	{{\bf{v}} _i} = {[{\cos} {\phi _i}{\sin}{\theta _i},{\sin}{\phi _i}{\sin}{\theta _i}]^T}.
	\end{equation}
	The distance between the central phase reference antenna and the ${\it{n}}$th antenna element of the ${\it{m}}$th layer can be expressed as 
	\begin{equation}\label{equ:distance_between_array}
	{\bf{q}} _{m,n} = {[{\cos}({\psi _{m,n}})md,{\sin}({\psi _{m,n}})md]^T},
	\end{equation}
	where ${\psi _{m,n}}{{ = n}}*\varphi $ denotes the polar angle of ${\it{n}}$th antenna element of ${\it{m}}$th antenna layer.
	The steering vector of signal ${{\it{s}}_{\it{i}}}$ with angle of arrival $({\phi _{\it{i}}},{\theta _{\it{i}}})$ can be expressed as
	\begin{equation}\label{equ:steering_vector}
	{{ {\bf{ a}}}_{\it i}}  = {[\rm{1},{\it{a}_{\rm{1,0}}}({\phi _{\it i}},{\theta _{\it i}}),...,{\it{a}_{{\rm{1},{\rm{2}^{\it b}} - \rm{1}}}}({\phi _{\it i}},{\theta _{\it i}}),...,{\it{a}_{\it{p} - \rm 1,{\rm{2}^{\it b}} - 1}}({\phi _{\it i}},{\theta _{\it i}})]^{\it{T}}},
	\end{equation}
	where $\rm{1}$ denotes that the phase of phase reference antenna element is zero and the remaining items in \eqref{equ:steering_vector} refer to the phase difference row vector composed of the phase difference between other antenna elements and PFA.
	
	If there are ${\it{l}}$ far field signals, then the steering matrix ${\bf{D}}$ can be described as
	\begin{equation}\label{equ:steering_matrix}
	{\bf{D}} = \left[ {{{\bf a}_1},{{\bf a}_2},...{{\bf a}_{\it i}}...,{{\bf a}_l}} \right],
	\end{equation}
	where ${{\bf a}_{\it i}}$ is the steering vector of the ${\it i}$th signal.
	
	Before beamforming, the received signal $ {{\bf x}}$ of SBSA can be expressed as
	\begin{equation}\label{equ:signal_before_beamforming}
	{{\bf x}}  = {\bf{D}} {{\bf s}}  + {{\bf n}},
	\end{equation}
	where ${{\bf s}}$ is the signal vector with the dimension of $\it{l} \times \rm{1}$. $\bf {{n}} $ is the additive gaussian white noise vector.
	
	After beamforming, the received signal $ {{y}}$ of SBSA can be expressed as
	\begin{equation}\label{equ:signal_after_beamforming}
	{{y}}  = {{\bf{w}}^H}{\bf x},
	\end{equation}
	where the weight vector of each antenna ${\bf{w}}$ can be expressed as
	\begin{equation}\label{equ:weight_vector}
	{{\bf{w}}} = [{{{w}}_{0,0}}{\rm{,}}{{{w}}_{1,0}},{{{w}}_{1,1}},...{{{w}}_{{\it {m,n}}}}...,{{{w}}_{\it p - \rm 1,{2^{\it b} - \rm 1}}}{{\rm{]}}^T},
	\end{equation}
	where ${{{w}}_{{\it {m,n}}}}$ is the weight of the $\it n$th antenna element of the $\it m$th layer.
	Then, the directional response of the antenna array after beamforming can be expressed as
	\begin{equation}\label{equ:directional_response}
	{{\bf{r}}}  = {{{\bf{w}}} ^H}{{\bf{D}}}.
	\end{equation}
	Based on linear constrained minimum variance (LCMV) criterion \cite{[LCMV]}, the beamforming problem can be described as
	\begin{equation}\label{equ:beamforming_problem}
	\begin{array}{l}
	\mathop {\min }\limits_{\bf{w}}  \qquad {{\bf{w}}^H}{\bf{R}}{\bf{w}},\\
	s.t\qquad  {{\bf{D}}^H}{\bf{w}} = {{\bf{r}} _d},
	\end{array}
	\end{equation}
	where ${{\bf{r}} _d}$ is the expected response vector of SBSA, ${\bf{R}}$ is the correlation matrix of $ {\bf{x}} $. By Lagrange multiplier algorithm, optimal weight vector ${\bf{w}}_{{\rm{opt}}}$ can be deduced as \cite{[LCMV]}
	\begin{equation}\label{equ:result_of_beamforming_problem}
	{{\bf{w}}_{{\rm{opt}}}} = {{\bf{R}}^{ - 1}}{\bf{D}}{({{\bf{D}}^H}{{\bf{R}}^{ - 1}}{\bf{D}})^{ - 1}}{{\bf{r}} _d}.
	\end{equation}
	The desired directional response of the antenna array ${{\bf{r}} _d}$ can be expressed as ${{\bf{r}} _d} = {{\bf{r}} _{ad}} \odot {{\bf{r}} _{pd}}$, where ${{\bf{r}} _{ad}}$ and ${{\bf{r}} _{pd}}$ denote desired amplitude directional response and desired phase directional response, respectively.
	
	In most case, ${{\bf{r}} _{pd}}$ is not considered while beamforming. But change of ${{\bf{r}} _{pd}}$ will influence the beamforming pattern performance. An efficient iterative beamforming algorithm based on two-step least-square method can be designed as algorithm \ref{alg:Framwork_1} \cite{[R_PD]}.

	\begin{algorithm}[htb]  
		\caption{Improved LCMV Beamforming Algorithm }  
		\label{alg:Framwork_1}  
		\begin{algorithmic}  
			\Require  \\
			Desired amplitude directional response: ${{\bf{r}} _{ad}}$. \\
			Initializes ${{\bf {w}}_{\it 0}} = {{\bf{R}}^{ - 1}}{\bf{D}}{({{\bf{D}}^H}{{\bf{R}}^{ - 1}}{\bf{D}})^{ - 1}}{{\bf{r}} _{ad}}$. \\
			The index of iterations: $i=0$. \\
			The threshold of iterations: $I_{max}$. 
			\Ensure  \\
			Final converged beamforming vector: ${ {\bf {w}}_{{\it{{\rm{opt}}}}}}$. \\
			
			\While{${{\bf {w}}_{\it i}}$ isn't convergent and ${\it i} \le I_{max}$}
			\State ${\it i} = {\it i} + 1$;
			\State ${{\bf{ r}}_{pd}} = {{\bf {w}}_{{\it i} - 1}}  {\bf{D}}  {\left[ {diag\left( {{{\bf{ r}}_{ad}}} \right)} \right]^{ - 1}}$;
			\State Normalize ${ {\bf{r}} _{pd}}$: ${\bar {\bf{r}} _{pd}} = \frac{{\bf{r}} _{pd}}{|{\bf{r}} _{pd}|}$;
			\State ${{\bf {w}}_{\it i}}{\rm{ = }} {{\bf{R}}^{ - 1}}{\bf{D}}{({{\bf{D}}^H}{{\bf{R}}^{ - 1}}{\bf{D}})^{ - 1}}{{\bf{r}} _{ad}} \odot { \bar {\bf{ r}}_{pd}}^H$;
			\EndWhile
			\State${ {\bf {w}}_{{{{\rm{opt}}}}}}{\rm{ = }}{{\bf {w}}_{\it i}}$.
		\end{algorithmic}  
	\end{algorithm}
	
	Algorithm \ref{alg:Framwork_1} can obtain the conventional beam with small sidelobe, but it cannot effectively adjust the beamwidth. To adjust the beam width, BABA is proposed in this paper. Details of BABA are introduced in Algorithm \ref{alg:Framwork_2}.
	
	\subsection{BABA Based on DBF}\label{BABA:DBF}
	
\begin{figure}[ht]
		\includegraphics[scale=0.55]{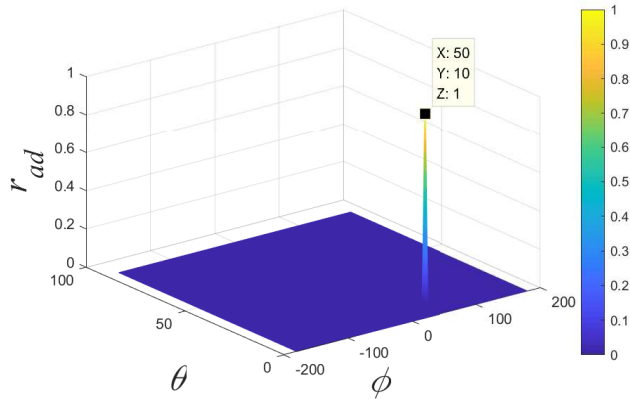}
		\centering
		\caption{Desired amplitude directional response in the direction of $({50^o},{10^o})$.}
		\label{fig:r_ad_1}
	\end{figure}

	Firstly, we define the desired amplitude directional response ${{\bf{ r}}_{ad}}$ as
	\begin{equation}\label{equ:r_ad}
	{{\bf{ r}}_{ad}}(\phi ,\theta ) = \frac{2}{{1 + {e^{(\frac{{{{(\theta  - {\theta _r})}^2}}}{{{{(\Delta \theta /2)}^2}}} + \frac{{{{(\phi  - {\phi _r})}^2}}}{{{{(\Delta \phi /2)}^2}}}) \times ln(2\sqrt 2  - 1)}}}},
	\end{equation}
	where $({\phi _{\it{r}}},{\theta _r})$ is desired signal direction, $(\Delta \phi ,\Delta \theta )$ denotes the expected beam azimuth and pitch width. The constant $ln(2\sqrt 2  - 1)$ is used to limit -3 dB half-power beamwidth. The desired amplitude directional response in the direction of $({50^{\rm{o}}},{10^{\rm{o}}})$ is shown in Fig. \ref{fig:r_ad_1}.
	
	Secondly, the desired beam angle range space $\it{A_{\it era}}$ is defined in BABA. The definition of $\it{A_{\it era}}$ is a rectangular area with desired direction $({\phi _{\it{r}}},{\theta _r})$ as the center and beam width $(\Delta \phi ,\Delta \theta )$ as the side length,
	\begin{equation}\label{equ:Aera_space}
	\begin{array}{l}
	{\it{A_{\it era} = \{ (}}\phi {\rm{,}}\theta {\rm{) | }}\\
	{\phi _r} - \frac{{\Delta \phi }}{2} \le \phi  \le {\phi _r} + \frac{{\Delta \phi }}{2}{\rm{,}}{\theta _r} - \frac{{\Delta \theta }}{2} \le \theta  \le {\theta _r} + \frac{{\Delta \theta }}{2}{\rm{\} }}.
	\end{array}
	\end{equation}
	
	Thirdly, the expected angle set ${{\it{C}}_{angle}}$ is obtained by uniformly dividing $\it{A_{\it era}}$ into $n$ subregions, which can be expressed as
	\begin{equation}\label{equ:C_angle}
	{{\it{C}}_{{\rm{angle}}}}{\rm{ = \{ (}}{\phi _1}{\rm{,}}{\theta _1}{\rm{),(}}{\phi _2}{\rm{,}}{\theta _2}{\rm{)}},...,{\rm{(}}{\phi _n}{\rm{,}}{\theta _n}{\rm{)\} }}.
	\end{equation}
	
	Then, the weight vector ${{\bf {w}}_{{\rm{opt}},{\it{i}}}}$ with the angle in ${{\rm{C}}_{angle}}$ as the expected direction can be generated by Algorithm \ref{alg:Framwork_1}. All weight vectors form a weight matrix
	\begin{equation}\label{equ:W_matrix}
	{\bf{W}{ = [}}{{\bf {w}}_{{{{\rm{opt}},1}}}}{\rm{,}}{{\bf {w}}_{{{{\rm{opt}},2}}}}{\rm{,}}...{\rm{,}}{{\bf {w}}_{{\rm{opt}},{\it{n}}}}{\rm{]}}.
	\end{equation}
	
	The fourth step is the key point of BABA. Based on convex optimization criterion, we construct a set of coefficients ${ {\bf{f}} _w}{\rm{ = [}}{{\it{f}}_1}{\rm{,}}{{\it{f}}_2}{\rm{,}}...{\rm{,}}{{\it{f}}_n}{\rm{]}}$ for ${\bf{W}}$. Based on MMSE algorithm, the beamforming problem can be described as follows
	\begin{equation}\label{equ:convex}
	\begin{array}{c}
	\mathop {\min }\limits_{{{ {\bf{f}} }_w}} (||{\bf r} - {{\bf r}_{ad}}|{|^2} + \beta ||{{\bf w} _{{\rm{opt}}}}|{|^2}),\\
	{\rm{s.t: }}{ {\bf{w}} _{{\rm{opt}}}} = {{\bf f} _w} {\bf W},\\
	\quad \quad  {\bf r} = { {\bf{w}}^H _{{\rm{opt}}}}{\bf D},
	\end{array}
	\end{equation}
	where {\bf D} is the steering matrix, the desired amplitude directional response ${{\bf r}_{ad}}$ can be obtained from \eqref{equ:r_ad}, the regularization factor  $\beta $ is designed for avoiding excessive growth of side lobe caused by large ${ {\bf{w}} _{{\rm{opt}}}}$. Details of BABA is shown in Algorithm \ref{alg:Framwork_2}.
	\begin{algorithm}[htb]  
		\caption{Beamwidth Adjustable Beamforming Algorithm }  
		\label{alg:Framwork_2}  
		\begin{algorithmic}  
			\Require  
			Expected direction: $({\phi _{\it{r}}},{\theta _r})$. \\
			\quad \quad Configuration of beam width: $(\Delta \phi ,\Delta \theta )$.
			\Ensure  
			Final converged beamforming vector: ${ {\bf{w}} _{{\rm{opt}}}}$. \\
			
			\State 1) Determine $\it{A_{\it era}}$;
			\State 2) Obtain ${{\it{C}}_{{\rm{angle}}}}$ by uniformly dividing $\it{A_{\it era}}$;
			\State 3) Obtain $\bf W$ by Algorithm \ref{alg:Framwork_1};
			\State 4) Obtain ${ {\bf{w}} _{{\rm{opt}}}}$ by solving problem \eqref{equ:convex}.
		\end{algorithmic}  
	\end{algorithm}
	In addition, \eqref{equ:convex} can be solved by CVX toolbox.
	
	\subsection{BABA Based on HBF}\label{BABA:HBF}
	
	\begin{figure}[ht]
		\includegraphics[scale=0.25]{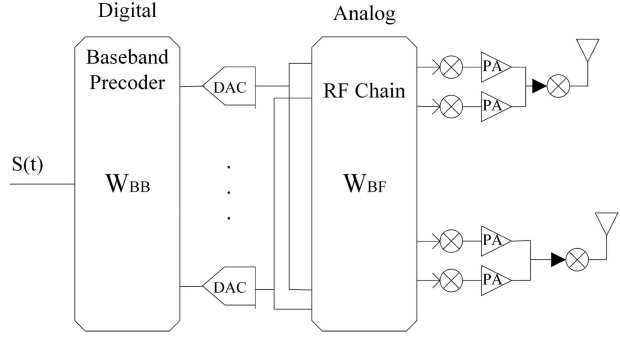}
		\centering
		\caption{Full-connected HBF structure.}
		\label{fig:HBF-full}
	\end{figure}
	
	Implementing digital beamforming in large-scale antenna systems presents many challenging issues: complexity, energy consumption, and cost. HBF is an important choice for practical large-scale antenna deployment \cite{[HBF]}. 
	
	As Fig. \ref{fig:HBF-full} shows, the digital signal from each transceiver can be delivered to any antenna after analog weighting. The directional response of the antenna array after HBF can be expressed as 
	\begin{equation} \label{equ:BABA-HBF-1}
	{\bf r} = {\bf W}_{BB} {\bf W}_{BF} {\bf D},
	\end{equation}
	where $\bf D$ is the steering matrix mentioned in section \ref{I-LCMV}, ${\bf W}_{BB}$ is the baseband combining matrix, ${\bf W}_{BF}$ is the  RF combining matrix. Then the beamforming problem can be described as 
	\begin{equation}\label{equ:BABA-HBF-2}
	\begin{array}{c}
	\begin{aligned}
	\mathop {\min }\limits_{{\bf W}_{BB},{\bf W}_{BF}} & ||{\bf r} - {{\bf r}_{ad}}|{|^2},\\
	{\rm{s.t: }} & {\bf W}_{BF} \in {\mathcal{F}}_{BF} , \\
	&||{\bf W}_{BB}{\bf W}_{BF}||^2_{F} = N_s,
	\end{aligned}
	\end{array}
	\end{equation}
	where ${{\bf r}_{ad}}$ is the desired amplitude directional response mentioned in section \ref{BABA:DBF}, $N_s$ is the number of data streams, ${\mathcal{F}}_{BF}$ is the set of feasible RF precoders, i.e., the set of RF matrices with constant-magnitude entries. A good reference on joint ${\bf W}_{BB}$ and ${\bf W}_{BF}$ design can be found in \cite{[HBF-full]}.

	\section{Sensing Capability of RSU}\label{sec:ROC}
	
	Receiver operating characteristic (ROC) denotes the relationship between the detection probability $p_d$ and false alarm probability $p_f$, which is addressed in this section.
	
	Based on section \ref{sec:CSM}, the echo signal of RSU after signal-level fusion can be expressed as
	\begin{equation}
	\left\{ {\begin{array}{*{20}{l}}
		{{{\rm{H}}_0}:}\\
		{{{\rm{H}}_1}:}
		\end{array}\begin{array}{*{20}{l}}
		{y = {\it{n}} + I}\\
		{y = P_{\rm{dfc}} \cdot \sum_{i} S_i + n + I}
		\end{array}}. \right.
	\label{equ:ROC_6}
	\end{equation}
	where ${\it{n}}$ denotes the additive noise that follows gaussian distribution ${\it{n}} \sim {{\mathcal{CN}}\rm{(0,}}\sigma _n^2{\rm{)}}$, the probability density function is
	\begin{equation}
	{\it{f}_n(y)} = \frac{1}{{\sqrt{2\pi}{\sigma_n}}}{\it{e}}^{( - \frac{y^2}{{2P_n}})},
	\label{equ:noise}
	\end{equation}
	with $P_n = \sigma^2_n$ being the noise power. The ground clutter $I$ follows Rayleigh distribution \cite{[rayleigh]},
	\begin{equation}\label{equ:rayleigh}
	{\it{f}_I(y)} =\left\{
	\begin{aligned}
	\frac{y}{\sigma^2_i}e^{-\frac{y^2}{2\sigma^2_i}}, & \quad y \ge 0 \\
	0, & \quad y \le 0 
	\end{aligned},
	\right.
	\end{equation}
	with $P_I = 2\sigma_i^2$ being the ground clutter power. The echo signal of each RSU ${\it{S}_i} = A{e^{j\varphi }}$, where $A$ and phase $\varphi$ are amplitude and phase respectively. According to \cite{[OFDM]}, the received echo power of RSU can be expressed as
	\begin{equation}
	{{\it{P}}_{\it{r,s}}} = \frac{{{P_t}{g_t}{g_{r}}{g_{p}}{\lambda ^2}\sigma }}{{{{(4\pi )}^3}{{{\it R}^4_{t,i}}}}}{\rm{ = }}{{||\it{S}_i}||_2}{{{ = {A}^2_i}}},
	\label{equ:radar_rec}
	\end{equation}
	then we have
	\begin{equation}
	{A} = \sqrt {\frac{{{P_t}{g_t}{g_{r}}{g_{p}}{\lambda ^2}\bar{\sigma} }}{{{{(4\pi )}^3}{{{R}^4_{t,i}}}}}},
	\label{equ:PD3}
	\end{equation}
	where $\it{R}_{t,i}$ is the distance between RSU and the sensing target, $\lambda$ is the wavelength of the echo, ${P_t}$ is the sending power, $g_p$ is the sensing processing gain, ${g_t}$ and ${g_{r}}$ are beam gain for sending and receiving, respectively. The sensing cross section $\bar{\sigma} = \pi r^2_s \rho$ with $\pi r^2_s$ being the area of sensing unit and $\rho \in (0,1)$ being the receiving echo probability.
	Since the area of sensing unit is affected by the width of the beam, the width of the beam also has an impact on the power of the echo signal and further affects the sensing capability of RSU, which will be analyzed by simulation in the following section. 
	
	Then, the probability density of $y$ can be derived as
	\begin{equation}
	{\begin{array}{*{20}{l}}
		{{\rm{p}}(y|{H_0}) = \frac{1}{{\sqrt{2\pi}{\sigma_n}}}{\it{e}}^{( - \frac{y^2}{{2\sigma_n^2}})}} \cdot {\it{f}_I(y)},\\
		{{\rm{p}}(y|{H_1}) = 
			\frac{1}{{\sqrt{2\pi}{\sigma_n}}}{\it{e}}^{( - \frac{(y-P_{\rm{dfc}} \cdot \sum_{i} A_i)^2}{{2\sigma_n^2}})}} \cdot {\it{f}_I(y)}.
		\end{array}}
	\label{equ:Radar_received_signal_1}
	\end{equation}
	The decision rule can be expressed as
	\begin{equation}
	\left\{ {\begin{array}{*{20}{l}}
		{{\rm{H_0:}}}\\
		{{\rm{H_1:}}}
		\end{array}\begin{array}{*{20}{l}}
		{y < \eta },\\
		{y > \eta },
		\end{array}} \right.
	\label{equ:Neyman-Pearson}
	\end{equation}
	where $\eta $ is the threshold of received signals' amplitude. The detection probability $p_d$ and false alarm probability $p_f$ can be further derived as
	\begin{equation}
	\begin{array}{c}
	{{\it{p}}_d} = \int\limits_\eta ^\infty  {  {p(y|{H_1})dy} }, \\
	{{\it{p}}_{\it{f}}} = \int\limits_\eta ^\infty  {  {p(y|{H_0})dy} }. 
	\end{array}
	\label{equ:PD_PF}
	\end{equation}
	Substitute \eqref{equ:rayleigh} and \eqref{equ:Radar_received_signal_1} into \eqref{equ:PD_PF}, $p_f$ and $p_d$ can be further derived as
	\begin{equation}
	{{\it{p}}_{\it{f}}} = \int\limits_\eta ^\infty \int\limits_0 ^\infty \frac{1}{{\sqrt{2\pi}{\sigma_n}}}{\it{e}}^{( - \frac{(z-y)^2}{{2\sigma_n^2}})} \cdot \frac{y}{\sigma^2_i}e^{-\frac{y^2}{2\sigma^2_i}}  dydz,
	\label{equ:PF}
	\end{equation}
	\begin{equation}
	{{\it{p}}_d} = \int\limits_\eta ^\infty \int\limits_0 ^\infty \frac{1}{{\sqrt{2\pi}{\sigma_n}}}{\it{e}}^{( - \frac{(z-y-P_{\rm{dfc}} \cdot \sum_{i} A_i)^2}{{2\sigma_n^2}})} \cdot \frac{y}{\sigma^2_i}e^{-\frac{y^2}{2\sigma^2_i}}  dydz.
	\label{equ:PD1}
	\end{equation}
	
	ROC can be deduced by \eqref{equ:PD3}, \eqref{equ:PF} and \eqref{equ:PD1}, which is difficult to directly calculate. Numerical integration is provided. However, when there is a big difference between the noise power $\sigma^2_n$ and the ground clutter power $\sigma^2_i$, the approximate relationship between ${{\it{p}}_d}$ and ${{\it{p}}_f}$ can be obtained.
	
	Case 1: $\sigma^2_n >> \sigma^2_i$, we ignore the ground clutter and get the formula for ${{\it{p}}_f}$ and ${{\it{p}}_d}$ as follows
	\begin{equation}
	\begin{aligned}
	{{\it{p}}_f} &=  \int\limits_\eta ^\infty \frac{1}{\sqrt{2\pi}\sigma_n}e^{\frac{y^2}{2\sigma^2_n}}dy 
	= Q(\frac{\eta}{\sigma_n}), \\
	{{\it{p}}_d} &=  \int\limits_\eta ^\infty \frac{1}{\sqrt{2\pi}\sigma_n}e^{\frac{(y-P_{\rm{dfc}} \cdot \sum_{i} A_i)^2}{2\sigma^2_n}}dy 
	= Q(\frac{\eta- P_{\rm{dfc}} \cdot \sum_{i} A_i}{\sigma_n}). \\
	\end{aligned}
	\label{equ:PF_PD_case1}
	\end{equation}
	Then, ROC can be deduced as
	\begin{equation}
	{{\it{p}}_d} = Q[\frac{\sigma_nQ^{-1}(\it{p}_f)-P_{\rm{dfc}} \cdot \sum_{i} A_i}{\sigma_n}].
	\label{equ:ROC_case1}
	\end{equation}
	
	Case 2: $\sigma^2_n << \sigma^2_i$, we ignore the gaussian white noise and get the formula for ${{\it{p}}_f}$ and ${{\it{p}}_d}$ as follows
	\begin{equation}
	\begin{aligned}
	{{\it{p}}_f} &=  \int\limits_\eta ^\infty \frac{y}{\sigma^2_i}e^{\frac{y^2}{2\sigma^2_i}}dy 
	= e^{(-\frac{\eta^2}{2\sigma^2_i})}, 
	\end{aligned}
	\label{equ:PF_case2}
	\end{equation}
	\begin{equation}
	{{\it{p}}_d} = \left\{
	\begin{aligned}
	\int\limits_\eta ^\infty  & F(y)dy
	= e^{[-\frac{(\eta-P_{\rm{dfc}} \cdot \sum_{i} A_i)^2}{2\sigma^2_i}]} &,\eta \ge P_{\rm{dfc}} \cdot \sum_{i} A_i\\
	& 1 &,\eta \le P_{\rm{dfc}} \cdot \sum_{i} A_i  
	\end{aligned}, \right.
	\label{equ:PD_case2}
	\end{equation}
	where 
	\begin{equation}
	F(y) = \frac{y-P_{\rm{dfc}} \cdot \sum_{i} A_i}{\sigma^2_i}e^{\frac{(y- P_{\rm{dfc}} \cdot \sum_{i} A_i)^2}{2\sigma^2_i}}.
	\end{equation}
	Then, ROC can be deduced as
	\begin{equation}
	{{\it{p}}_d} = \left\{
	\begin{aligned}
	&e^{-\frac{[\sqrt{-2\sigma^2_iIn({{\rm{p}}_f})}-P_{\rm{dfc}} \cdot \sum_{i} A_i]^2}{2\sigma^2_i}} &,{{\it{p}}_f} \le e^{-\frac{(P_{\rm{dfc}} \cdot \sum_{i} A_i)^2}{2\sigma^2_i}}\\
	&1 &,{{\it{p}}_f} \ge e^{-\frac{(P_{\rm{dfc}} \cdot \sum_{i} A_i)^2}{2\sigma^2_i}}  
	\end{aligned}. \right.
	\label{equ:ROC_case2}
	\end{equation}
	The reasonability of the above two approximations is verified by simulation in section \ref{sec:Radar-Performance}-C.

	\section{Simulation Results}\label{sec:Radar-Performance}
	
	Simulation parameters used in this section are shown in table \ref{Parameter:simulation} \cite{[OFDM]} \cite{[parameters_1]} \cite{[parameters_3]}.
	
	\begin{table}[t]
		\caption{Simulation parameters adopted in this paper.}
		\label{Parameter:simulation}
		\begin{tabular}{l|l|l|l|l|l}
			\hline
			\hline
			Items & Value & Meaning of the parameter & Items & Value & Meaning of the parameter \\ \hline
			\it{h} & 10 m & The height of antenna array & $f_c$ & 24 GHz & Carrier frequency of the echo \\ \hline
			$\lambda$ & 0.0125 m & Wavelength of the echo & ${P_{t}} {{\rm{g}}_{t}}$ & 20 dBm & ISM limit 20 dBm \\ \hline
			$\it{k}$ & $1.38*{10^{ - 23}}$ J/K & Boltzmann constant & ${T_{\rm{abs}}}$ & 290 K & Absolute temperature \\ \hline
			$P_n$ & $-94$ dBm & Thermal noise $\it{k}T_{\rm{abs}}B$ for 100 MHz & $P_I$ & $(-110\sim-70)$ dBm & Ground clutter power \\ \hline
			${F_n}$ & 6 dB & \begin{tabular}[c] {@{}l@{}} Identical system noise figure \end{tabular} & $\rho $ & 0.1 & Receiving echo probability \\ \hline
			$g_{p}$ & 54.2 dB \cite{[OFDM]}& Sensing processing gain & $g_{r}$ & various & Beam gain for receiving \\ \hline
			${\it{p}}$ & 32+1 & Number of antenna layers & ${2^b}$ & 32 & Number of antennas per layer \\ \hline
			$({\phi _{\it{r}}},{\theta _r})$ & $({60^o},{45^o})$ & Expected direction of sensing beam & $(\Delta \phi ,\Delta \theta )$ & $({12^o},{6^o})$ & Width of sensing beam\\ \hline
			$v$ & $20 \sim 40$ km/h & Speed of vehicles \\ \hline
		\end{tabular}
	\end{table}
	
	\subsection{Simulation of BABA}
	
\begin{figure}[ht]
		\centering
		\subfigure{\includegraphics[width=0.6\textwidth]{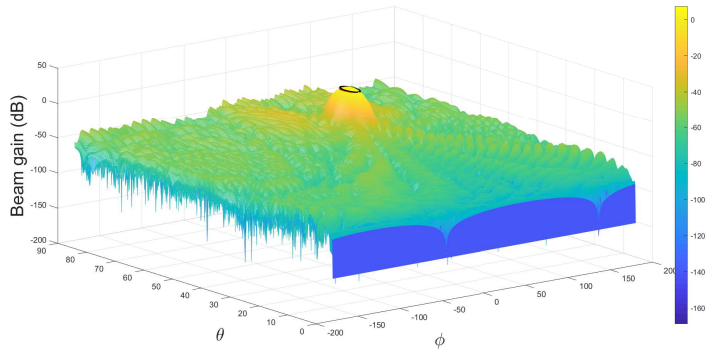}%
			\label{linkSuccessProb_K10}}
		\hfil
		
		\caption{Beamforming gain with direction of $({60^o},{45^o})$ and beamwidth of $({12^o},{6^o})$.}
		\label{fig:perfect_1}
	\end{figure}

	Configurations of the antenna array and sensing beam are shown in table \ref{Parameter:simulation}. Beamforming gain with direction of $({60^o},{45^o})$ and beamwidth of $({12^o},{6^o})$ is shown in Fig. \ref{fig:perfect_1}.
	
	According to \cite{[ant_1]}, the -3dB beam width is adopted as the metric of beam width. The red and black area in Fig. \ref{fig:perfect_1} represents the 3 dB attenuation region of the main beam in simulation and ideal conditions, respectively. To analyze the performance of BABA, an parameter $e_b$ is defined as
	\begin{equation}\label{equ:error_1}
	{{{e}}_b} = \frac{{{E_{area}}}}{{{B_{area}}}},
	\end{equation}
	where the beamwidth area ${{B_{area}}}$ is an elliptic region with pitch and azimuth width as major semiaxis and minor semiaxis respectively. The error area ${{E_{area}}}$ can be defined as
	\begin{equation}\label{equ:error_2}
	\begin{array}{c}
	{{\it{E}}_{area}} = \{ (\theta ,\phi ){\rm{|}}(\theta ,\phi ) \in {B_{area}}{\it{\& r(}}\theta ,\phi {\rm{) < max(}} {\bf{r}} {\rm{) - 3}}\} \\
	\quad \quad \quad...
	U\{ (\theta ,\phi ){\rm{|}}(\theta ,\phi ) \notin {B_{area}}{\it{\& r(}}\theta ,\phi {\rm{) > max(}} {\bf{r}} {\rm{) - 3}}\}.
	\end{array}
	\end{equation}	
	According to \eqref{equ:error_2}, ${{E_{area}}}$ consists of two regions: the region in ${{B_{area}}}$ where the beam attenuation is greater than 3dB, and the region outside ${{B_{area}}}$ where the beam attenuation is less than 3dB.
	
	\begin{figure}[ht]
		\includegraphics[scale=0.55]{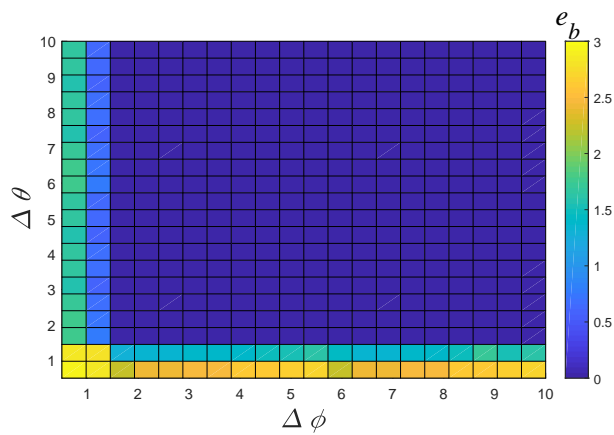}
		\centering
		\caption{${{\it{e}}_b}$ of various beam width with direction of $({60^o},{45^o})$.}
		\label{fig:error_of_various_width}
	\end{figure}
	
	\begin{figure}[ht]
		\includegraphics[scale=0.55]{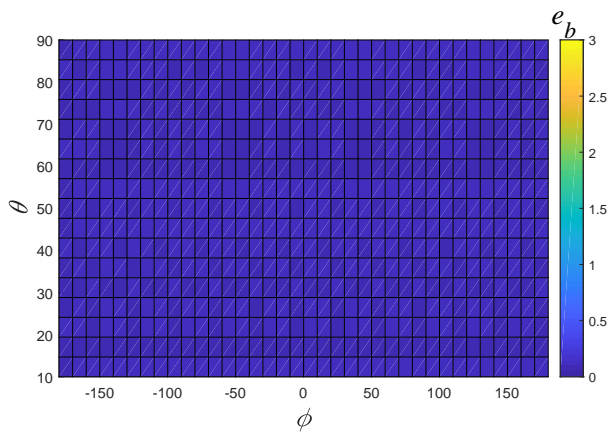}
		\centering
		\caption{${{\it{e}}_b}$ of various direction with beamwidth of $({6^o},{3^o})$.}
		\label{fig:error_of_various_angle}
	\end{figure}
			
\begin{figure}[ht]
		\includegraphics[scale=0.5]{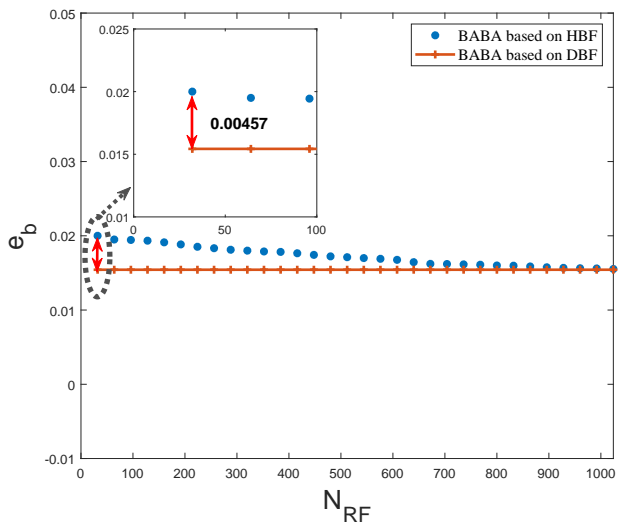}
		\centering
		\caption{$e_b$ of BABA based on HBF and DBF with direction of $(60^o,45^o)$ and beamwidth of $(6^o,3^o)$.}
		\label{fig:HBF_1}
	\end{figure}

	As Fig. \ref{fig:error_of_various_angle} shows, ${{{e}}_b}$ of various directions with beamwidth of $({6^o},{3^o})$ is smaller than $0.1$, which means that the beam direction has little influence on the performance of BABA. 
	The relationship between ${{{e}}_b}$ and beamwidth is shown in Fig. \ref{fig:error_of_various_width}. When the width of beam is smaller than ${1^o}$, ${{{e}}_b}$ is larger than $1$, which means BABA can not generate a beam with width smaller than ${1^o}$.
	Under the same conditions, the angular resolution of the beam formed by the antenna array is limited by the number of antenna arrays. In other words, the minimum beam width that can be generated by the antenna array is limited by the number of antenna arrays.
	BABA, the beamforming algorithm proposed in this paper, can obtain the beam with adjustable width through the idea of weighting optimization, but cannot decrease the minimum beam width that can be generated by the antenna array. 
	Therefore, $e_b$ is relatively large when the required beam width is less than the angular resolution that the antenna array itself can achieve. On the contrary, when the required beam width gradually increases, we can get the beam with the required width by means of weighting optimization, and $e_b$ is relatively small. Therefore, as the beamwidth increases, $e_b$ turns to be reduced. 
	Simulation shows that the minimum beam width that the antenna array used in this paper can generate is about ${1^o}$. Therefore, when the required beam width is less than ${1^o}$, $e_b$ is relatively large, as shown in Fig. \ref{fig:error_of_various_width}. The performance can be improved by increasing the number of antenna array elements.
	\textsc{}
	As Fig. \ref{fig:HBF_1} shows, when the expected direction is $(60^o,45^o)$ and the expected beam width is $(6^o,3^o)$, ${{{e}}_b}$ of BABA based on HBF is bigger than DBF. With the number of RF chains increases, ${{{e}}_b}$ decreases. Moreover, Fig. \ref{fig:HBF_1} verifies that similar performance can be achieved by using HBF and DBF schemes. 
	{
	In the urban traffic environment, the speeds of vehicles are relatively slow. To achieve high-quality sensing at all times, RSU needs to generate sensing beams pointing in different directions with high performance. As Fig. \ref{fig:error_of_various_angle} and Fig. \ref{fig:error_of_various_width} shows, BABA proposed in this paper can generate high performance sensing beams pointing in different directions when the beam width meets certain conditions, which verifies that BABA is suitable for urban traffic environment.}
	
	\subsection{Simulation of Spatial Registration}
	
	\subsubsection{Simulation scenario} 
	
\begin{figure}[ht]
		\includegraphics[scale=0.23]{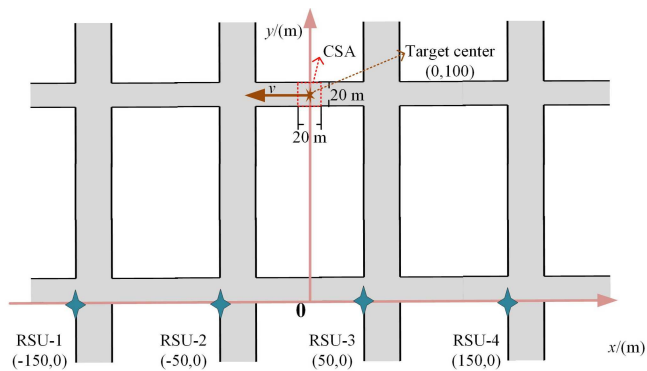}
		\centering
		\caption{Simulation scenario of AB-SRA.}
		\label{fig:simulation_scene}
	\end{figure}

	As Fig. \ref{fig:simulation_scene} shows, four RSUs operate in the same coordinate system. They are deployed at (-150,0), (-50 ,0), (50,0) and (150,0) separately. The height of RSU is 10 m. CSA is a rectangular area whose vertexes are located at (10,90), (10,110), (-10,90) and (-10,110) separately. RSU forms a sensing unit on CSA by adjusting the direction and width of sensing beam. The sensing unit needs to scan the whole CSA to complete the cooperative sensing of CSA. The radius of the sensing unit is marked as $r_s$. 
	{
	The speeds of vehicles in urban traffic environments are mostly in the range of $20 \sim 40$ km/h and the width of CSA is 20 m. Thus, vehicles run in the CSA for roughly $1.8 \sim 3.6$ s. Since the main computational operation of AB-SRA is in the generation of the sensing beam, to achieve high-quality sensing for vehicles, the direction and width of the sensing beam generated by RSU need to achieve fast adjustment. 
	Based on the simulation parameters in Table \ref{Parameter:simulation}, the time for realizing BABA using MATLAB's cvx toolbox is about 1 s.
	Therefore, the AB-SRA and BABA proposed in this paper are suitable for urban traffic environment. 
	For higher speed traffic environments, the target tracking problem needs to be considered, which will be discussed further in Section \ref{sec:Conclusion}.
}

	\subsubsection{Simulation results} 
	
\begin{figure}[ht]
		\includegraphics[scale=0.55]{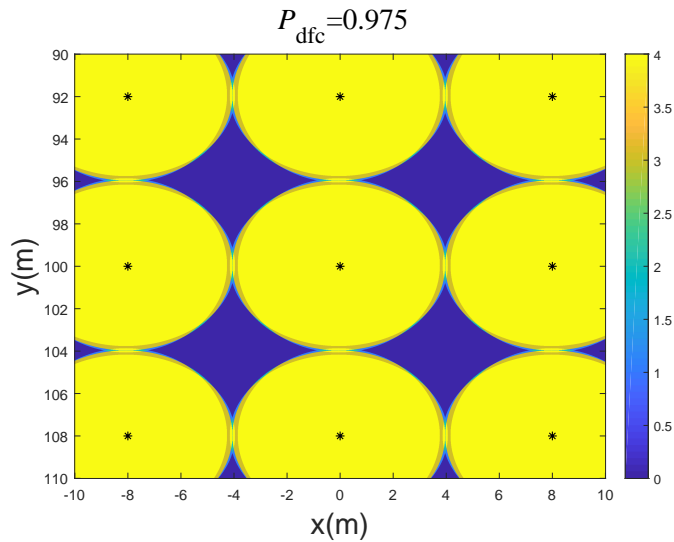}
		\centering
		\caption{Power distribution after noncoherent accumulation with $r_s=4$ m.}
		\label{fig:SP_1}
	\end{figure}

\begin{figure}[ht]
			\includegraphics[scale=0.55]{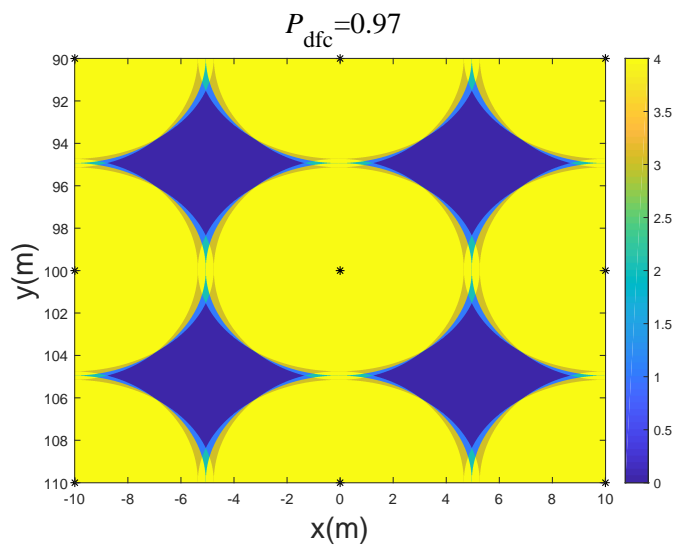}
			\centering
			\caption{Power distribution after noncoherent accumulation with $r_s=5$ m.}
			\label{fig:SP_2}
		\end{figure}

	Different the tracking-level and decision-level fusion, the aim of spatial registration in signal-level fusion is to make the sensing units of each RSU coincide.
	The more accurate the matching of sensing units of different RSUs is, the higher the power of target echo signal is, and the higher the detection probability of target is. The power distribution can reflect the echo signal power of the target in the sensing unit, and further reflect the detection probability of the target. Since the probability of target detection is one of the important indicators of target sensing, it is reasonable to adopt the power distribution as the spatial registration performance indicator of signal level fusion.
	According to \cite{[signal_fusion_space_reg]}, the performance of spatial registration algorithm can be judged by $P_{\rm{dfc}}$. 
	$P_{\rm{dfc}}$ represents the matching degree of the sensing units of different RSUs. 
	Assuming that the area of the sensing units of $N$ RSUs are $S_i,{i=1,2,…,N}$ and the area of the overlapping part is $S_o$, then $P_{\rm{dfc}}$ can be expressed as  
	\begin{equation}
	P_{\rm{dfc}} =  \frac{(N \cdot S_o)}{\sum_(i=1)^N {S_i}}.
	\end{equation}
	If spatial registration is realized perfectly, $P_{\rm{dfc}}=1$. 
	Fig. \ref{fig:SP_1} and Fig. \ref{fig:SP_2} shows power distribution after noncoherent accumulation with parameter $r_s=4$ m and $r_s=5$ m, respectively. The black star symbol is the center of the sensing unit and the color distribution indicates the statistical distribution of RSU energies after a simple noncoherent signal-level fusion. 

\begin{figure}[ht]
		\includegraphics[scale=0.5]{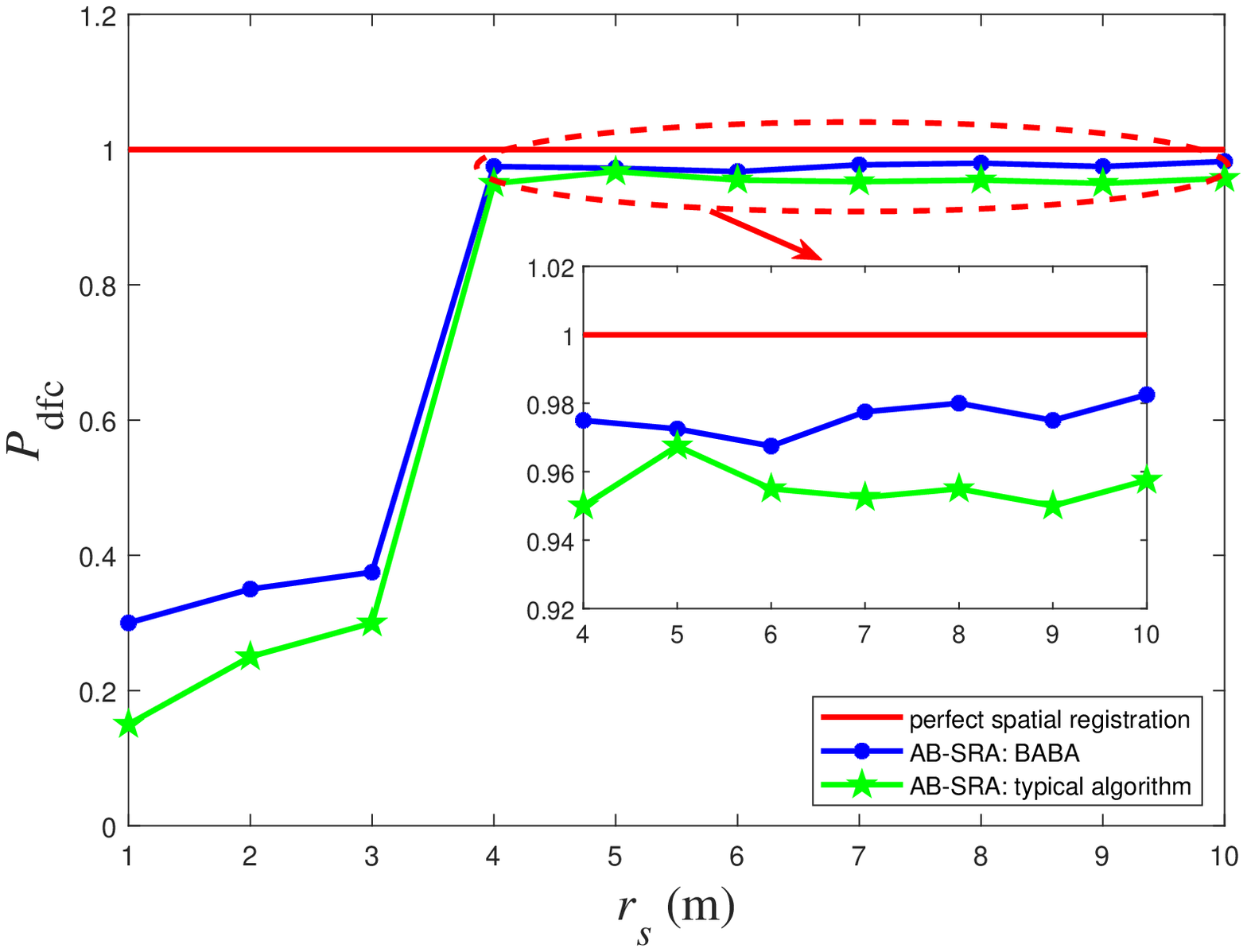}
		\centering
		\caption{the relationship between $P_{\rm{dfc}}$ and $r_s$.}
		\label{fig:P_dfc}
	\end{figure}

	According to Fig. \ref{fig:error_of_various_width} and Fig. \ref{fig:error_of_various_angle}, the performance of BABA measured by $e_b$ varies with the change of the width and direction of sensing beam. The size of the sensing unit is related to the width and direction of sensing beam. Therefore, $e_b$ varies with the radius of the sensing unit $r_s$. Since the performance of AB-SRA is limited by BABA, $P_{\rm{dfc}}$ is different for different radius of the sensing unit $r_s$, as shown in Fig. \ref{fig:P_dfc}. When $r_s \le 3$, $P_{\rm{dfc}}$ is much smaller than 1, which means that AB-SRA has poor performance. This is because that BABA has poor performance on generating a beam with a beamwidth smaller than $1^o$. When $r_s>3$, $P_{\rm{dfc}}$ is close to 1, which means AB-SRA has good performance. 
	
	In this paper, AB-SRA with typical width adjustable beamforming algorithm is compared with AB-SRA with BABA proposed in this paper. As shown in Fig. \ref{fig:P_dfc}, AB-SRA based on BABA proposed in this paper has slightly better overall performance than AB-SRA based on typical beamforming algorithms \cite{[TA_1], [TA_2]}.
	
	According to \cite{[signal_fusion_space_reg]}, the power distribution is different with different relaxation coefficient $m$. Different $P_{\rm{dfc}}$ can be obtained with different $m$, just as shown in table \ref{signal_fusion_space_reg}. 
	It can be found that with the decrease of M, the larger $P_{\rm{dfc}}$ is, the better spatial registration performance is.  When m=1/2, $P_{\rm{dfc}}$ can reach 0.952, which is close to the perfect spatial registration algorithm.  However, according to \cite{[signal_fusion_space_reg]}, as $m$ decreases, the time of cooperative sensing for the same area also increases. 
	Since the algorithm proposed in \cite{[signal_fusion_space_reg]} and AB-SRA adopt different ideas to solve the spatial registration problem, it is difficult to make an intuitive comparison between them in the same graph. 
	In order to obtain better spatial registration performance, a smaller $m$ is needed, and thus a longer sensing time is consumed. However, AB-SRA proposed in this paper does not have the problem of consuming more sensing time. As shown in Fig. \ref{fig:P_dfc}, when $r_s>3$, $P_{\rm{dfc}}$ almost tends to be stable, and the spatial registration performance is close to the perfect spatial registration. The disadvantage of AB-SRA is that need to sacrifice the angular resolution of sensing to a certain extent, which has been analyzed previously. 
	In practical application, we can choose an appropriate spatial registration algorithm according to the specific requirements of the scene. If the requirement of the angle resolution of sensing is not high but that of the sensing time is high, AB-SRA can be used as a priority. Otherwise, the spatial registration algorithm in \cite{[signal_fusion_space_reg]} can be considered.
		
	\begin{table}[ht]
		\caption{$P_{\rm{dfc}}$ with different relaxation coefficient $m$.}
		\label{signal_fusion_space_reg}
		\begin{tabular}{llllllllllllllllc|cccc}
			\cline{17-21} 
			\cline{17-21}
		&  &  &  &	&  &  &  & &  &  &  &	&  &  &  & m           & 1/2  & 2/3  & 3/4  & 1  \\ \cline{17-21} 
		&  &  &  &	&  &  &  & &  &  &  &	&  &  &  & $P_{\rm{dfc}}$ & 0.952 & 0.904 & 0.753 & 0.459 \\ \cline{17-21} 
		\end{tabular}
	\end{table}
	
	\subsection{Simulation of Sensing Capability}

\begin{figure}[ht]
		\includegraphics[scale=0.5]{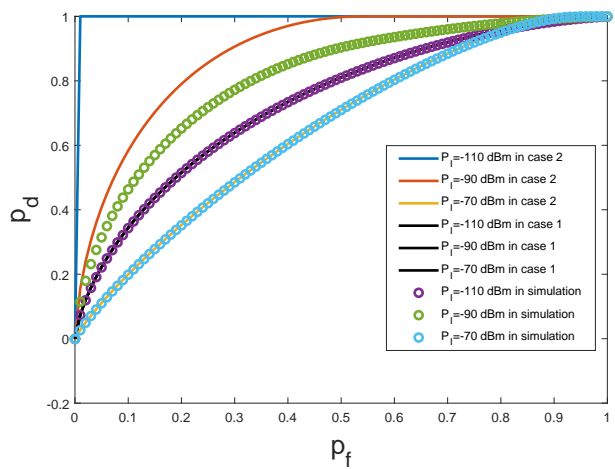}
		\centering
		\caption{ROC for single RSU with different ground clutter power $P_I$.}
		\label{fig:SBS_ROC}
	\end{figure}

	ROC denotes the relationship between detection probability $p_d$ and false alarm probability $p_f$. With a given $p_f$, the bigger $p_d$ is, the better sensing capability is. Assuming that $r_s=4m$, $\rho=0.1$, Fig. \ref{fig:SBS_ROC} shows ROC for single RSU with different ground clutter power $P_I$ under the condition of the noise power $P_n=-94$ dBm. As Fig. \ref{fig:SBS_ROC} shows, when $P_I=-70$ dBm that much larger than $P_n$, ROC obtained by simulation is close to ROC derived from case 2 in section \ref{sec:ROC}; when $P_I=-110$ dBm that much smaller than $P_n$, ROC obtained by simulation is close to ROC derived from case 1 in section \ref{sec:ROC}. The simulation results show that the above two approximate schemes are reasonable to some extent.
	
	\begin{figure}[ht]
		\includegraphics[scale=0.5]{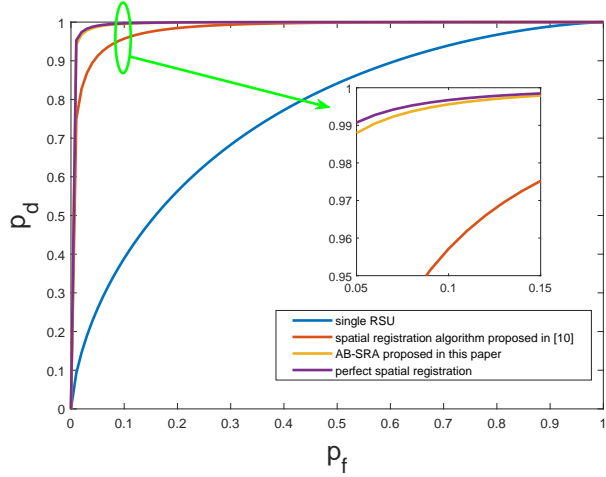}
		\centering
		\caption{ROC for RSU-CRSN with ground clutter power $P_I = -110$ dBm.}
		\label{fig:ROC}
	\end{figure}
	
	Fig. \ref{fig:ROC} shows ROC for RSU-CRSN with ground clutter power $P_I = -110$ dBm.
	As Fig. \ref{fig:ROC} shows, the detection capability of RSU-CRSN is better than single RSU. When $p_f$ is 0.1, $p_d$ of single RSU is 0.3891, and $p_d$ of RSU-CRSN based on perfect spatial registration is 0.9967, which means that the sensing capability of RSU-CRSN has been greatly improved. Moreover, ROC of RSU-CRSN based on perfect spatial registration and AB-SRA are similar, which means that the performance of AB-SRA is close to perfect spatial registration. Compared with spatial registration algorithm proposed in \cite{[signal_fusion_space_reg]}, when $p_f =0.1$, $p_d$ of RSU-CRSN based on AB-SRA is 0.0385 larger, which means AB-SRA is closer to perfect spatial registration than spatial registration algorithm proposed in \cite{[signal_fusion_space_reg]}.
	
	\subsection{Sensing Capability at Different Beamwidth}
	According to \eqref{equ:PD3}, the amplitude of echo signal is different for different sensing cross section. Since the area of sensing unit is affected by the width of the beam, the width of the beam also has an impact on the amplitude of the echo signal and further affects the sensing capability of RSU.
	
	Assuming that $P_I = -110$ dBm and $P_n = -94$ dBm, which means that $\sigma^2_n \gg \sigma^2_i$, then ROC can be expressed as \eqref{equ:ROC_case1}. 
	According to \cite{[3D_beam]}, the receiving beam gain can be approximated as 
	\begin{equation}\label{equ:3D_beam_1}
	g_r = \frac{26000}{\Delta \theta \cdot \Delta \phi }.
	\end{equation}
	Therefore, $g_r$ decreases with increasing beamwidth, but $\bar \sigma = \pi r^2_s \rho$ increases with increasing beamwidth. 
	Substitute \eqref{equ:PD3} and \eqref{equ:3D_beam_1} into \eqref{equ:ROC_case1}, ROC can be expressed as
	\begin{equation}\label{equ:ROC_rs}
	{{\it{p}}_d} = Q[\frac{\sigma_nQ^{-1}(\it{p}_f)-P_{\rm{dfc}} \cdot{ \sum_{i} }\sqrt {\frac{{{\rm{26000}}{P_t}{g_t}{g_{p}}{\lambda ^{\rm{2}}}\pi r^{\rm{2}}_s \rho }}{{{{\Delta \theta_i \cdot \Delta \phi_i}{({\rm{4}} \pi )}^{\rm{3}}}{{{R}^{\rm{4}}_{t,i}}}}}}}{\sigma_n}].
	\end{equation}
	
	Based on the simulation scenario mentioned above, $R_{t,i}$ can be obtained as follow.
	\begin{table}[ht]
		\caption{$R_{t,i}$ for different RSUs.}
		\label{Parameter:R_t}
		\begin{tabular}{lllc|cccc}
			\cline{4-8} 
			\cline{4-8}
			&  &  & RSU           & RSU-1  & RSU-2  & RSU-3  & RSU-4  \\ \cline{4-8} 
			&  &  & $R_{t,i}$ (m) & 180.56 & 112.25 & 112.25 & 180.56 \\ \cline{4-8} 
		\end{tabular}
	\end{table}
	
\begin{figure}[ht]
		\includegraphics[scale=0.5]{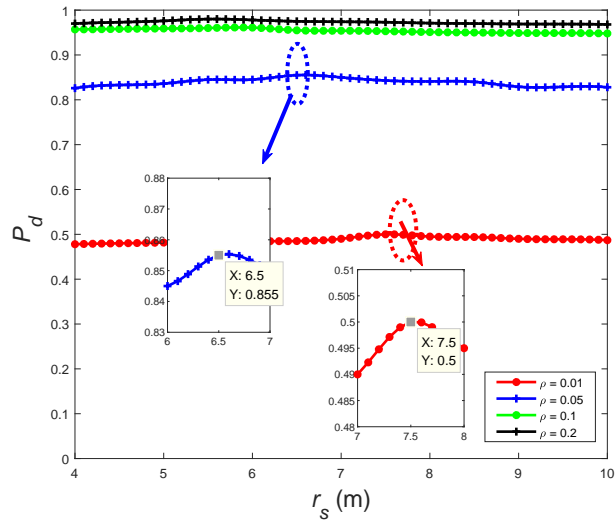}
		\centering
		\caption{$p_d$ of $p_f = 0.1$ with different $r_s$.}
		\label{fig:PD_r}
	\end{figure}

	$r_s$ is used to uniformly represent the beam width of each RSU. Consider Model 1 mentioned in section \ref{subsec:registration-algorithm_Specific_Scenarios}, beamwidth $\Delta \theta_i$ and $ \Delta \phi_i$ can be obtained by \eqref{equ:limitation_1} and \eqref{equ:limitation_2}.
	As Fig. \ref{fig:PD_r} shows, when $p_f=0.1$ and $\rho = 0.05$, $p_d$ increases first with the increase of $r_s$, and after $r_s \ge 6.5$ m , $p_d$ decreases with the increase of $r_s$. When $r_s \approx 6.5$ m, $p_d$ is the biggest, which means the sensing capability is the best. Moreover, the optimal $r_s$ decreases with the increase of $\rho$.  
	
	\section{Conclusion and Discussion}\label{sec:Conclusion}
	
	In this paper, we propose a multiple RSUs cooperative radar sensing network with signal-level fusion that can assist vehicles to sense the traffic environment. To achieve spatial registration, we present AB-SRA. By adjusting the width of the sensing beam, we can make the sensing area of each RSU coincide. To adjust the beamwidth flexibly, BABA is proposed in this paper. This paper focuses on the improvement of spatial registration algorithm on RSU-CRSN's sensing capability. The spatial registration algorithm can help RSU-CRSN obtain higher power gain $P_{\rm{dfc}}$, thus obtaining higher power echo signal, thus obtaining better sensing capability. Simulation results show that BABA works well when the beam width is greater than $1^o$. Moreover, simulation results of the RSU-CRSN's performance show that RSU-CRSN has significant promotion in term of sensing capability.
	
	{
	In a highly mobile traffic environment, the sensing beam needs to follow the high-speed moving targets at all times to achieve high-quality sensing, which means that the direction and width of the sensing beam need to achieve fast adjustment. This will lead to a new problem, the target tracking problem.
	Thus, how to achieve fast direction switching of the sensing beam is the main factors affecting spatial registration.
	Simulation results show that BABA proposed in this paper can generate high performance sensing beams pointing in different directions when the beam width meets certain conditions, but BABA requires weighting multiple beamforming results to achieve the purpose of adjustable sensing beam width, so the computational complexity is relatively high.
	Reducing the desired beam angle range space $A_{era}$ can reduce the complexity of BABA to a certain extent, but it also leads to a reduction in the beamforming performance.
	Therefore, it is important to reduce the computational complexity of the beamforming algorithm to generate the desired sensing beam in a shorter time to achieve real-time tracking of moving targets. 
	One solution to reduce the computational complexity without sacrificing the sensing performance is that 
	RSU firstly generates the codebook composed of ${\bf w}_{opt}$ of the sensing beam with different width and direction, and when implementing the tracking vehicle, directly obtains ${\bf w}_{opt}$ by codebook, and quickly generate the desired sensing beam. But this also increases the storage complexity to some extent.
	This factor will be the focus of our future work for the design of spatial registration in the highly mobile traffic environment.
}

	\bibliographystyle{IEEEtran} 
	\bibliography{reference}	
	
	\begin{IEEEbiography}[{\includegraphics[width=1in,height=1.25in,clip,keepaspectratio]{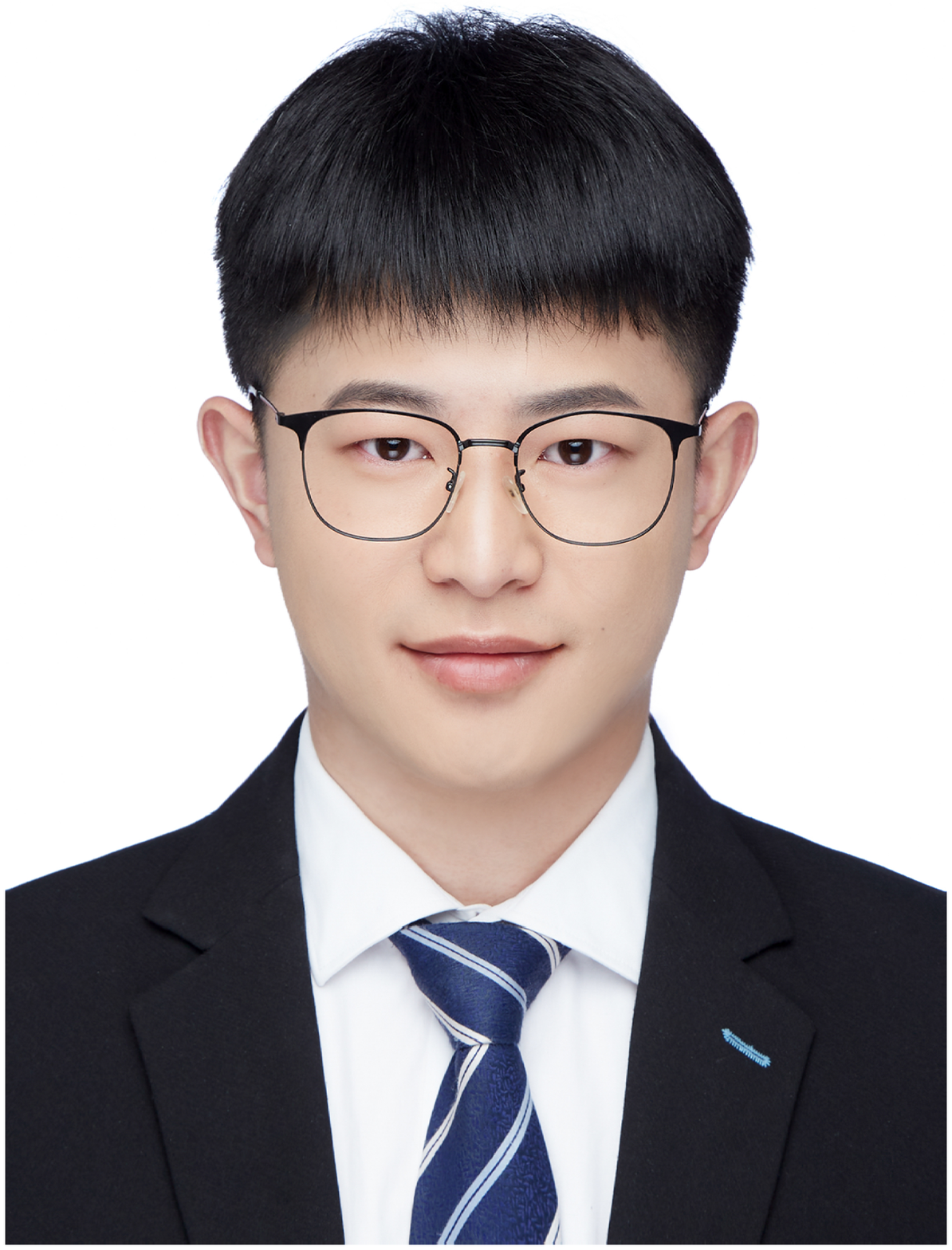}}]{Wangjun Jiang}
	 received the B.S. degree in School of Electronic and Information Engineering, Beijing Jiaotong University (BJTU) in 2019. He is currently pursuing her Ph.D. degree with Beijing University of Posts and Telecommunication (BUPT). His research interests include integrated sensing and communication and network sensing.
	\end{IEEEbiography}
	
	\begin{IEEEbiography}[{\includegraphics[width=1.1in,height=1.4in,clip,keepaspectratio]{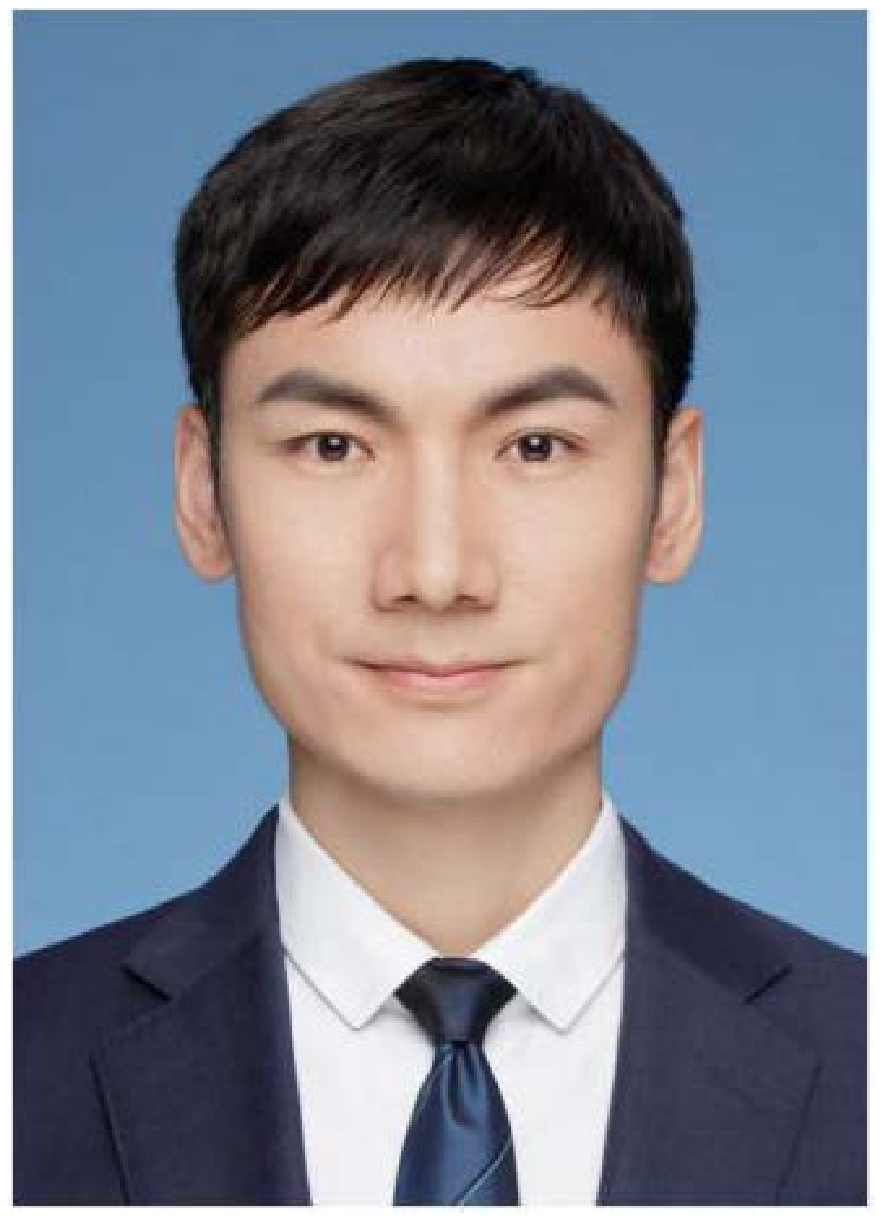}}]
		{Zhiqing Wei}
		(S'12-M'15) received his B.E. and Ph.D. degrees from BUPT in 2010 and 2015. Now he is an associate professor at BUPT. He was granted the Exemplary Reviewer of IEEE Wireless Communications Letters in 2017, the Best Paper Award of International Conference on Wireless Communications and Signal Processing 2018. He was the Registration Co-Chair of IEEE/CIC International Conference on Communications in China (ICCC) 2018 and the publication Co-Chair of IEEE/CIC ICCC 2019. His research interest is the performance analysis and optimization of mobile ad hoc networks.
	\end{IEEEbiography}
	
	\begin{IEEEbiography}[{\includegraphics[width=1.1in,height=1.4in,clip,keepaspectratio]{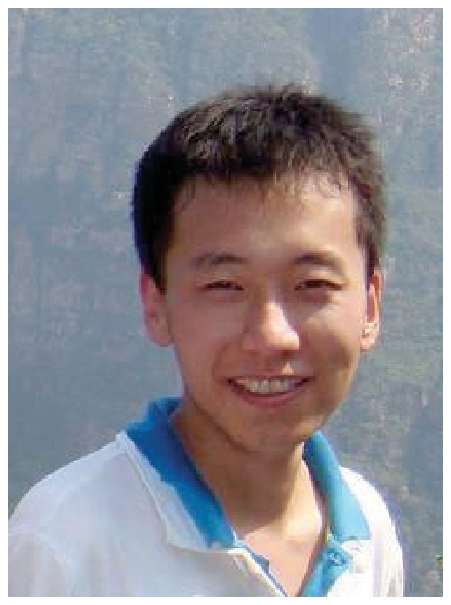}}]{Bin Li}
		received the bachelor’s degree in electrical information engineering from the Beijing University of Chemical Technology in 2007, and the Ph.D. degree in communication and information engineering from the Beijing University of Posts and Telecommunications (BUPT) in 2013. In 2013, he joined BUPT, where he is currently an Associate Professor with the School of Information and Communication Engineering. His current research interests include signal processing for wireless communications and machine learning, such as millimeter-wave communications, UAV communications, MIMO communication/radar systems. He received the 2011 ChinaCom Best Paper Award, the 2015 IEEE WCSP Best Paper Award.  
	\end{IEEEbiography}
	
	\begin{IEEEbiography}[{\includegraphics[width=1.1in,height=1.4in,clip,keepaspectratio]{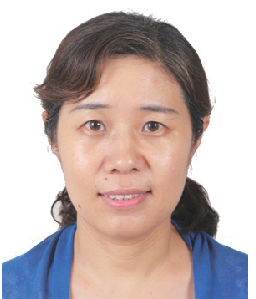}}]{Zhiyong Feng}
		(M’08-SM’15) received her B.S., M.S., and Ph.D. degrees from BUPT, Beijing, China. She is a Professor with the School of Information and Communication Engineering, BUPT, and the director of the Key Laboratory of Universal Wireless Communications, Ministry of Education, China. Her research interests include wireless network architecture design and radio resource management in 5th generation mobile networks (5G), spectrum sensing and dynamic spectrum management in cognitive wireless networks, universal signal detection and identification, and network information theory. She is a senior member of IEEE and active in standards development, such as ITU-R WP5A/5C/5D, IEEE 1900, ETSI, and CCSA.
	\end{IEEEbiography}
	
	\begin{IEEEbiography}[{\includegraphics[width=1.1in,height=1.4in,clip,keepaspectratio]{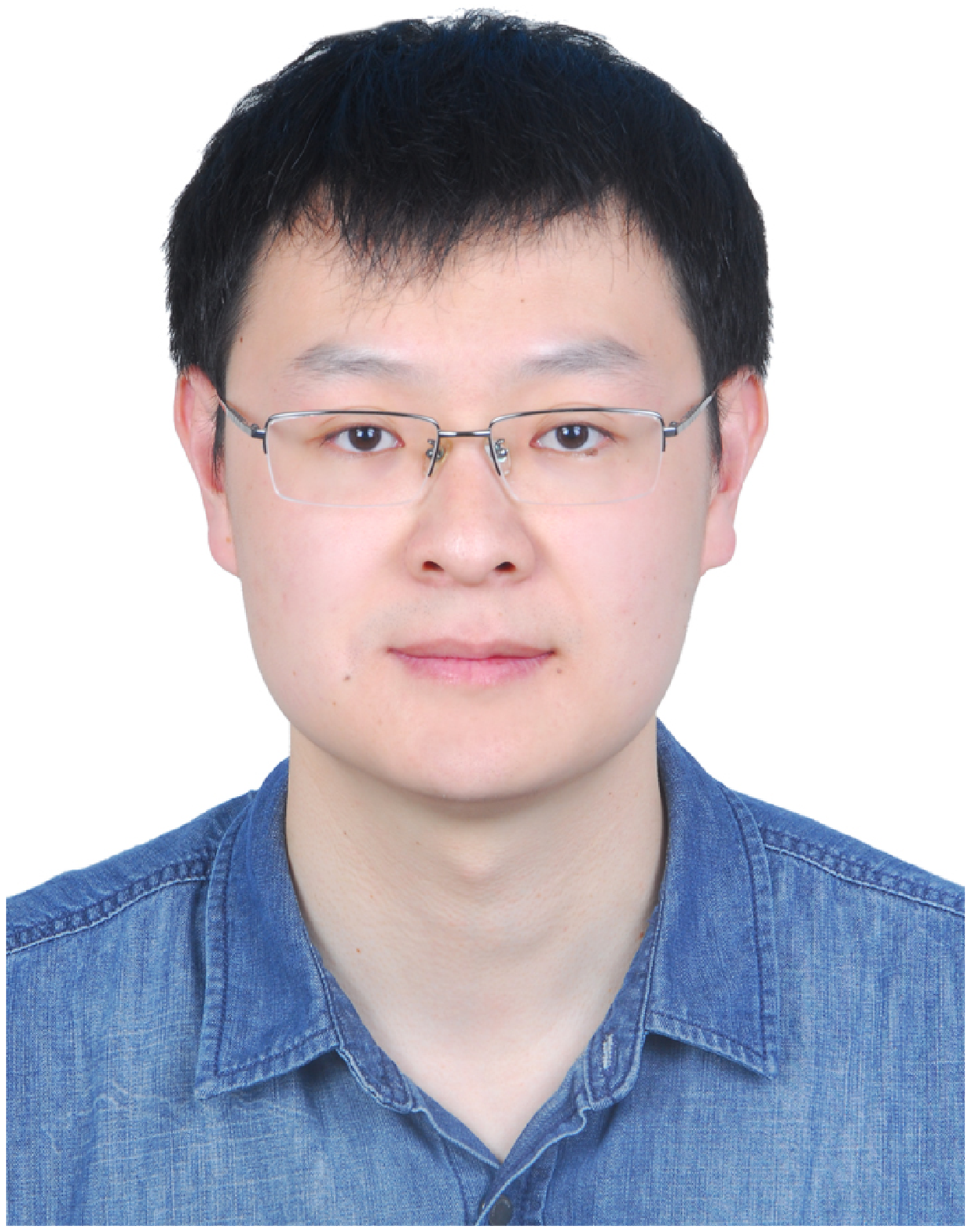}}]{Zixi Fang}
		received his B.E. degree from Guilin University of Electronic Technology, Guilin, China. He received his M.E. degree from North China Electric Power University (NCEPU), Baoding, China. He is currently
		pursuing the Ph.D. degree with the School of Information and Communication Engineering, Beijing University of Posts and Telecommunications (BUPT), Beijing, China. His research interests
		include wireless communication theory, joint radar and communication, etc.
	\end{IEEEbiography}

	\ifCLASSOPTIONcaptionsoff
	\newpage
	\fi

\end{document}